\documentclass[twocolumn,tightenlines,superscriptaddress,longbibliography]{revtex4-1}
\usepackage{amssymb,amsthm,amsmath,enumerate,url, natbib, graphicx, float, csvsimple, todonotes, enumitem,booktabs}
\usepackage{hyperref}
\hypersetup{
    colorlinks=true,       
    linkcolor=black,          
    citecolor=orange,        
    filecolor=magenta,      
    urlcolor=cyan           
}

\newcommand{\bs}[1]{\boldsymbol{#1}}

\newcommand{\reals}{\mathbb{R}}

\newcommand{\expect}[1]{\ensuremath{\left\langle#1\right\rangle}}

\begin{document}
\author{Travis L. Scholten}
\affiliation{IBM T.J. Watson Research Center, Yorktown Heights, NY 10598, USA}
\author{Yi-Kai Liu}
\affiliation{National Institute of Standards and Technology, Gaithersburg, Maryland 20899, USA}
\affiliation{Joint Center for Quantum Information and Computer Science, University of Maryland, College Park, Maryland 20742, USA}
\author{Kevin Young}
\affiliation{Quantum Performance Lab, Sandia National Laboratories, Livermore, California 94550, USA}
\author{Robin Blume-Kohout}
\affiliation{Quantum Performance Lab and Center for Computing Research, Sandia National Laboratories, Albuquerque, New Mexico 87185, USA}
\affiliation{Center for Quantum Information and Control, University of New Mexico, Albuquerque, New Mexico 87131, USA}

\title{Classifying single-qubit noise using machine learning}

\begin{abstract}
Quantum characterization, validation, and verification (QCVV) techniques are used to probe, characterize, diagnose, and detect errors in quantum information processors (QIPs). An important component of any QCVV protocol is a mapping from experimental data to an estimate of a property of a QIP. Machine learning (ML) algorithms can help automate the development of QCVV protocols, creating such maps by learning them from training data. We identify the critical components of ``machine-learned" QCVV techniques, and present a rubric for developing them. To demonstrate this approach, we focus on the problem of determining whether noise affecting a single qubit is coherent or stochastic (incoherent) using the data sets originally proposed for gate set tomography. We leverage known ML algorithms to train a classifier distinguishing these two kinds of noise. The accuracy of the classifier depends on how well it can approximate the ``natural" geometry of the training data. We find GST data sets generated by a noisy qubit can reliably be separated by linear surfaces, although feature engineering can be necessary. We also show the classifier learned by a support vector machine (SVM) is robust under finite-sample noise.
\end{abstract}
\date{\today}
\maketitle

Characterizing the errors and imperfections in a quantum information processor (QIP) is necessary to understand what is going wrong with it, to identify ways to fix the errors, and to improve the processor's performance. To do so, quantum characterization, validation, and verification (QCVV) techniques are used. There are many QCVV techniques, in part because QIPs have many \emph{properties} to be characterized at various levels of detail. (A ``property" being anything that describes the processor's behavior.)  Interesting, relevant properties are quite diverse: they can be binary (e.g., the presence of leakage), real-valued (e.g., $T_{1}$ and $T_{2}$ times, logical error rates, average fidelities), matrix-valued (e.g., process matrices), or fall within some other category.

\begin{figure}[h]
\includegraphics[width=\columnwidth]{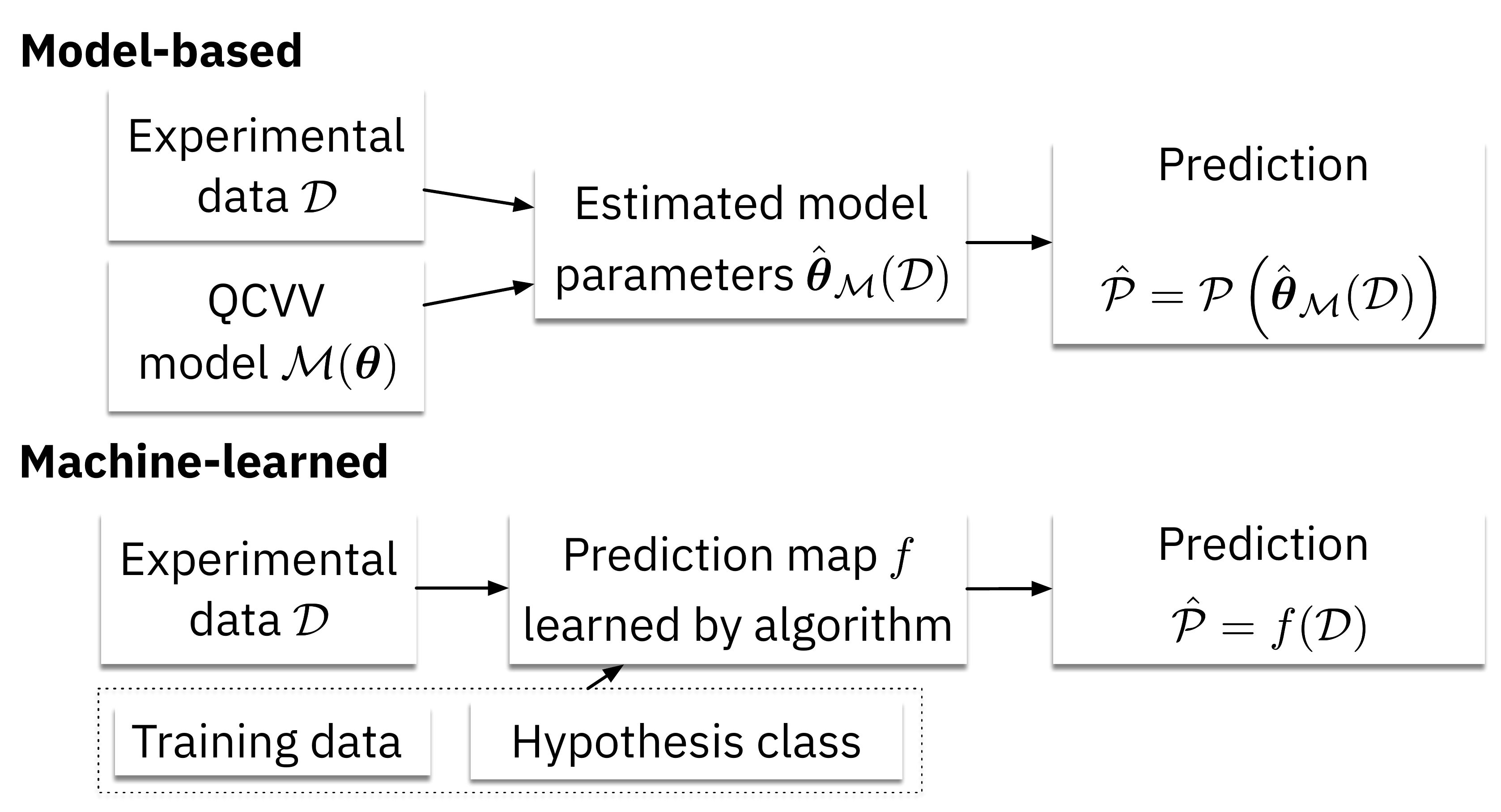}
 \caption{\textbf{Comparing ``model-based" and ``machine-learned" QCVV}. QCVV techniques use an experimental data set $\mathcal{D}$ to make an inference about some property $\mathcal{P}$ of a QIP. ``Model-based" QCVV techniques make predictions based on the estimated parameters of a statistical model $\mathcal{M}(\bs{\theta})$. In the NISQ era, developing new QCVV techniques within this paradigm can be challenging. We introduce and discuss ``machine-learned" QCVV as an alternative paradigm in which predictions are made using a prediction map learned by a machine learning algorithm.}
\label{fig:overview}
\end{figure}

Many QCVV techniques fall into a ``model-based" paradigm: they use a \emph{statistical model}, which is a parameterized family of probability distributions over experimental data. Estimates of these parameters can be used to predict important properties of the QIP. (See Figure \ref{fig:overview}.) They can be used to diagnose problems, mitigate them, and improve the correctness of future results obtained from the (improved) QIP. Until recently, the number of parameters to be estimated has been manageable, largely because QIPs contained just a few qubits.

As QIPs grow more sophisticated, characterizing them becomes harder.  Large, multi-qubit QIPs have more properties that need to be characterized, and inventing a new  QCVV technique to probe a new property of interest is a nontrivial task. New methods for developing QCVV techniques will be required to keep pace with the development of QIPs. This is especially true in the ``noisy, intermediate-scale quantum" (NISQ) era \cite{Preskill2018}, as novel multi-qubit processors are being developed and deployed by IBM, Google, IonQ, Rigetti Quantum Computing, and others \cite{Kelly2018, Vu2017, Wright2019, Zeng2017}.

However, NISQ processors will have more qubits, and are likely to demand new QCVV techniques.  Many existing techniques demand resources -- e.g., number of experimental configurations, repetitions of each experiment, or classical postprocessing -- that grow rapidly with the number of qubits \cite{Blume-Kohout2013a}. The techniques that \emph{do} scale well usually describe the QIP's behavior in a coarse-grained way, by using a statistical model with a modest number of parameters \cite{Magesan2012a, Magesan2011}. These techniques don't enable detailed characterization of novel types of noise expected to appear in next-generation QIPs, such as crosstalk or correlated errors.  Finally, new techniques will be necessary to probe the ``holistic" performance of QIPs on tasks stressing the entire processor \cite{Cross,Blume-Kohout2019}.

But inventing a new QCVV technique is hard.  It typically demands \emph{creativity}, \emph{effort}, and \emph{time}.  New QCVV techniques are informed by significant domain-specific expertise, and distilling the complex behavior of a QIP into a meaningful set of characterizable properties requires thoughtful effort. In this paper, we ask ``\emph{Can expertise be augmented by computation, to let scientists develop QCVV techniques more rapidly?}''  We suggest the answer is ``yes,'' by examining how machine learning (ML)  \cite{Samuel1959, Hastie2008, Goodfellow2016, James2017} can automate one of the more challenging aspects of creating a QCVV technique. We call such a QCVV technique created with the help of ML ``machine-learned".

\section{Overview: machine-learned QCVV techniques}
\label{sec:intro}

\subsection{Advantages and disadvantages of using machine learning (ML) for QCVV tasks}
\label{subsec:ml_helps}

Any QCVV technique requires an \emph{experiment design} (``What kind of data will be collected?") and a \emph{data processing pipeline} (``How will that data be processed?"). In principle, ML algorithms can help to generate both. In this paper, we focus on data processing, and on using ML algorithms to learn good maps from data to inferred properties \footnote{In the conclusions (Section \ref{sec:conclusions}) we discuss how ML could also be used to construct experiment designs.}. We assume an expert has already specified an experiment design that produces data ($\mathcal{D}$) from which the property of interest ($\mathcal{P}$) can be inferred.  With the experiment design specified, solving the characterization task ``Infer $\mathcal{P}$ from $\mathcal{D}$." requires finding a prediction map $f: \mathcal{D}\rightarrow \mathcal{P}$ that yields the right prediction with high probability. 

Most existing QCVV techniques use prediction maps designed using statistical theory. The designer posits a parameterized statistical model $\mathcal{M}(\bs{\theta})$ that, for each value of the parameters $\bs{\theta}$, predicts the probability distribution of observed data, or a coarse-graining of it \footnote{For example, the model in randomized benchmarking predicts certain averages, rather than individual circuit probabilities}.  Then, the designer chooses a statistical estimation procedure (e.g. maximum likelihood or Bayesian inference \cite{Wasserman2004}) that maps the data $\mathcal{D}$ to an estimate $\hat{\bs{\theta}}_{\mathcal{M}}(\mathcal{D})$ of the model parameters describing the QIP that generated $\mathcal{D}$.  Finally, the prediction for the property $\mathcal{P}$ is inferred as its value for the estimated parameters: $\hat{\mathcal{P}} = P(\hat{\bs{\theta}}_{\mathcal{M}}(\mathcal{D}))$. 

For example, if $\bs{\theta}$ is a process matrix describing a gate, and the property of interest is ``Is the gate's error coherent?'', then the inferred answer is ``yes'' if $\hat{\bs{\theta}}$ corresponds to an undesired unitary rotation.  This approach is illustrated in the top part of Figure \ref{fig:overview}.

Such a QCVV technique is only as good as the models at its heart. There is no unique, obvious way to choose a good model -- George Box famously observed ``All models are wrong; some are useful.''\cite{Box1978}. A model with too many parameters will produce inaccurate estimates and falsely detect effects.  But overly simple models, insufficiently rich to capture the QIP's behavior and fit the data, will produce suspect or biased inferences about $\mathcal{P}$. Perhaps most importantly, a model capturing the desired property has to exist or be invented by a creative theorist before statistical machinery can be deployed.

A ``machine-learned" QCVV protocol need not rely on a statistical model. Its core is the prediction map $f$ learned by an ML algorithm. ``Learning" means identifying, out of a large set of candidate maps (the algorithm's \emph{hypothesis class}), a map that (1) works reliably on training data, and (2) satisfies some robustness criterion that suggests this reliability will extend to future, as-yet-unseen data. Machine-learned QCVV is viable in part because some ML algorithms are known to be particularly good at learning approximations to functions.  So \emph{if} a good approximation to the relationship between $\mathcal{D}$ and $\mathcal{P}$ exists -- e.g., one that could be derived via statistical theory -- then some ML algorithm should be able to find it \footnote{Some ML algorithms are in fact \emph{universal function approximators} \cite{Debao1993}, and can approximate arbitrarily well any given function (subject to mild regularity conditions).}.

Both machine-learned and model-based QCVV protocols use data, but whereas the model-based approach uses data to estimate parameters of a statistical model, machine learning cuts out the middle part, and uses data to choose $f$ directly. Because ML algorithms infer an analysis map \emph{inductively}, by generalizing from a large array of sample cases (training data, either synthetic or real), there is no obligation to come up with a good statistical model. This shifts the burden of effort from statistics (which requires reasoning about hypothetical data that might be observed) to collecting large amounts of training data.

In this work, we think of the prediction map learned by ML algorithms as a \emph{decision surface} separating data sets coming from QIPs with different values of the property of interest. Then, the hypothesis class of the algorithm is the set of all possible decision surfaces it could learn. Each decision surface gives rise to a decision rule, which is the approximation to $f$. As we will show, the interplay between the geometry of the decision surfaces within the algorithm's hypothesis class and the geometry of the data used to train the algorithm affects the quality of the predictions for the property.

Going from model-based to machine-learned QCVV techniques does come with several costs:
\begin{enumerate}
\item The prediction function learned by the algorithm may not make sense to humans, and not lend itself to insight or the development of intuition. This is particularly true for powerful algorithms such as neural networks \cite{McCulloch1943, Farley1954, Hinton2006}.
\item Generating sufficient training data can require lots of experiments and/or large amounts of simulation time. Creating synthetic data might become extremely costly once NISQ processors become (essentially) unsimulable on classical computers.
\item If the hypothesis class of the ML algorithm is not rich enough, then the prediction map learned by the algorithm will generally perform poorly. On the other hand, if the hypothesis class is extremely rich, efficiently training the algorithm may be hard. Choosing a good hypothesis class can be just as tricky as choosing a good model.
\item Many ML algorithms have user-specified \emph{hyperparameters} that control their behavior and affect the accuracy of the decision surface they learn. Hyperparameter tuning, like statistical model selection, is often necessary to maximize accuracy and minimize overfitting. This tuning can be expensive.
\end{enumerate}  

Replacing statistical inference with machine learning merely trades one set of challenges for another. The challenges of machine-learned QCVV are somewhat different from those of model-based QCVV, and thus machine-learned QCVV may perform well in situations where model-based QCVV does not.

\subsection{Problem statement and key results}

We investigate the use of supervised classification algorithms to learn a high-accuracy classifier for determining whether the noise affecting a single qubit is coherent or stochastic. This is a special case of a previously-studied problem, \emph{estimating the coherence of noise}, for which there are known QCVV techniques \cite{Wallman2015, Feng2016}. Our intention here is not to demonstrate a better solution, but to show that ML algorithms \emph{can} automatically produce solutions to this problem.

We first outline a rubric for developing a machine-learned QCVV technique based on supervised learning algorithms (Section  \ref{sec:ml_qcvv_framework}). We use this rubric  to show ``off-the-shelf" supervised classification algorithms can be trained to classify between coherent and stochastic noise. We find that the \emph{geometry} of QCVV datasets, plus the algorithm's hyperparameters, governs the accuracy of the decision surface it learns (Sections \ref{sec:application} and \ref{sec:clf_gst}). We also find that \emph{linear} classifiers -- ones that learn separating hyperplanes -- can have comparable performance to nonlinear classifiers, but only if \emph{feature engineering} is used to change the geometry of the QCVV datasets (Section \ref{subsec:feature_engineering_separable}). Finally, we examine the robustness of the support vector machine in the presence of finite-sample noise (Section \ref{sec:finite_sampling}). We conclude with an outlook on the viability of machine-learned QCVV for NISQ processors (Section \ref{sec:conclusions}).

\subsection{Related work}
\label{subsec:ml_for_qcvv}

The work reported here lies at the intersection of machine learning and quantum computing, a subfield in its own right called ``quantum machine learning" (QML) \cite{Schuld2015,Biamonte2017, Go2018, Dunjko2018}. Research in QML can be divided into 3 parts: machine learning algorithms that would be run on fault-tolerant hardware \cite{Harrow2009, Kerenidis2016}, ``quantum-inspired" algorithms that de-quantize a quantum algorithm to yield a new classical algorithm \cite{Tang2019, Tang2018, Gilyen2018}, and using classical ML algorithms to characterize quantum systems \cite{Huembeli2019,VanNieuwenburg2017,Carrasquilla2017,ChNg2017,Wetzel2017,Zhang2018,Granade2012,Youssry2019,Gupta2017,Magesan2015,Iten,Torlai2018}.

We consider data generated by performing measurements on a QIP. That measurement record (i.e., the experimental data) is classical in nature. We will use classical ML algorithms for processing that data. Consequently, this work falls into 3rd category (``classical ML for quantum systems").

\section{Using ML to generate new QCVV techniques}
\label{sec:ml_qcvv_framework}

\begin{figure}
\includegraphics[width=\columnwidth]{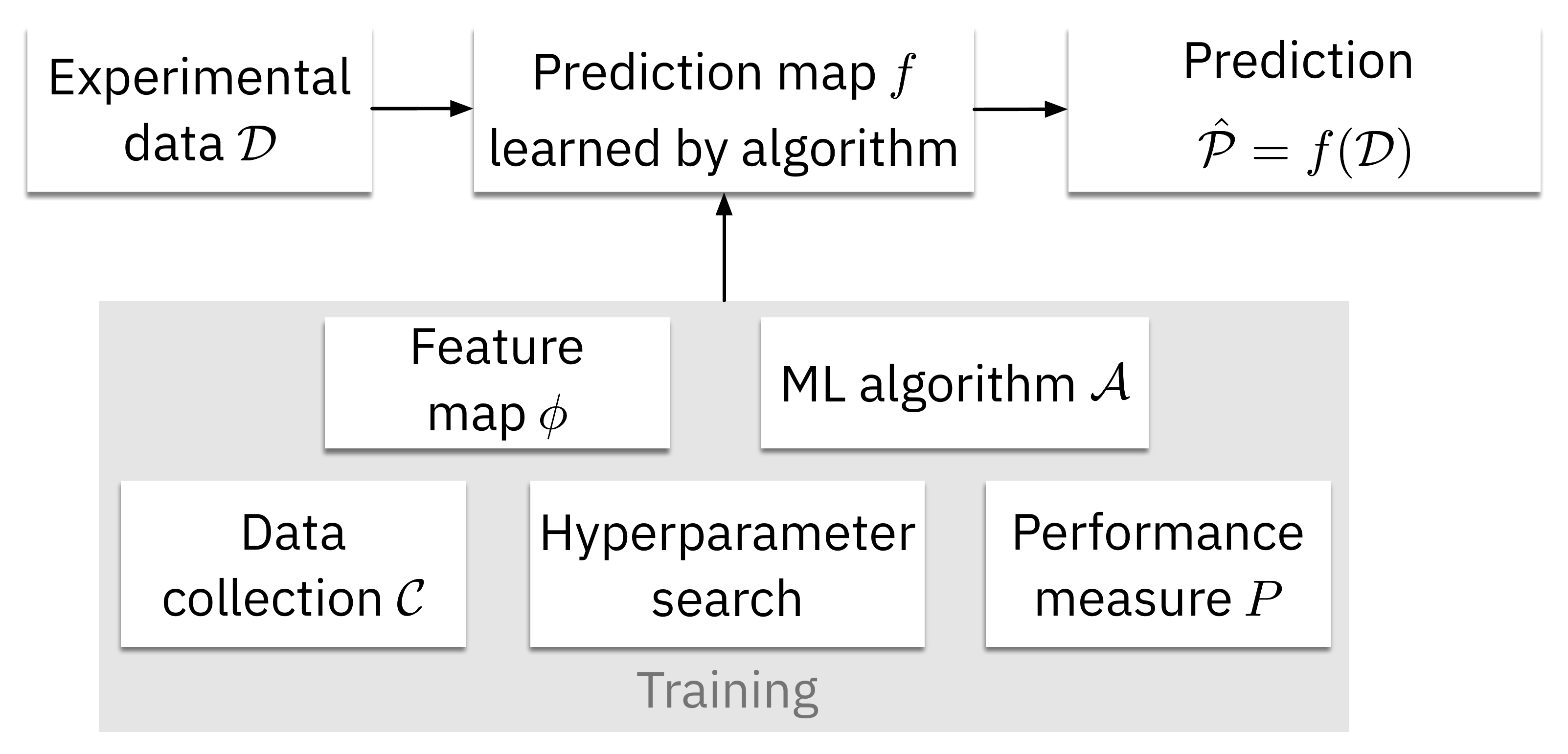}
 \caption{\textbf{Structure of machine-learned QCVV techniques}. Using ML algorithms to learn a QCVV technique requires different components from traditional, model-based QCVV techniques. In particular, the statistical model $\mathcal{M}(\bs{\theta})$ is replaced by an ML algorithm $\mathcal{A}$ that learns over a \emph{hypothesis class} of candidate functions. This learning is done using training data taken from a data collection $\mathcal{C}$. After training, the algorithm outputs a prediction map $f:\mathcal{D}\to\mathcal{P}$.}
\label{fig:ml_framework}
\end{figure}

There are several ways ML algorithms could be used to create a QCVV technique. Here, we lay out one particular rubric -- which we believe to be somewhat general, but not universal -- for using \emph{supervised learning} to do so \footnote{Supervised learning algorithms learn a map from data to property using data sets with known properties, whereas algorithms for unsupervised learning \emph{discover} structure within data.  The canonical QCVV task is to estimate a specific property, so supervised learning is more relevant.}. In this rubric, there are 7 general components, 5 of which are depicted in the ``Training" box in Figure \ref{fig:ml_framework}.

\subsubsection{The property of interest, $\mathcal{P}$ }
The property $\mathcal{P}$ is the answer to the question ``What do we want to know about the QIP?" In supervised learning, the property is used to label the training data: for each value of $\mathcal{P}$, some data is generated, and the ML algorithm has access to both that data and its associated label. This allows the algorithm to recognize what data are ``typically'' generated by a QIP with that value of the property. In this work, we focus on a binary property of interest. In general, other types of properties can be learned as well: for example, supervised regression algorithms could be used to estimate error rates or other continuous properties.

\subsubsection{An experiment design, defining experimental data $\mathcal{D}$}
The experiment design defines which experiments will be run to produce a dataset $\mathcal{D}$, and optionally the order in which they are run. Many QCVV experiments can be described by quantum circuits; the QIP performs these circuits, and generates outcomes.   So ``data" means ``The outcomes that result from running all the circuits in the experiment design."  If the QIP's behavior is assumed to be stationary, then $\mathcal{D}$ can simply consist of the aggregated outcome frequencies of the circuits \footnote{Except for Section \ref{sec:finite_sampling}, we consider the infinite-sample limit, so that the outcome frequencies are the outcome probabilities.}. Otherwise $\mathcal{D}$ comprises the time-stamped outcome (``click") for each repetition (``shot") of a each circuit.

\subsubsection{A feature map $\phi$ for embedding data into a feature space}
Making QCVV data amenable to analysis by ML algorithms requires embedding it into a \emph{feature space}, $\mathcal{F}$, which is typically isomorphic to $\reals^n$ for some $n$.  The embedding is described by a feature map, $\phi$, mapping each QCVV data set $\mathcal{D}$  to a \emph{feature vector} $\mathbf{f}$ (i.e., $\phi: \mathcal{D} \rightarrow \mathbf{f} \in \mathcal{F}$).

If ML algorithms do not perform very well on data embedded using a given $\phi$, there are at least 2 ways to improve performance by modifying it: feature engineering and feature selection. 

\emph{Feature engineering} augments $\mathcal{F}$ with new features derived by applying functions to existing features.  Feature engineering changes the geometric structure of the data, and can make properties learnable by the algorithm. Feature engineering helps ML algorithms that are underfitting the data to perform better by giving them new features to learn on. Feature engineering can be done explicitly or implicitly (e.g., using kernel methods.) We use and discuss feature engineering later (see Section \ref{subsec:feature_engineering}).

\emph{Feature selection} removes redundant and unnecessary features by determining which features are useful or necessary for a given task and disregarding others.  Feature selection helps ML algorithms avoid overfitting the data. We do not consider feature selection here. 

\subsubsection{A data processing pipeline, centered around an ML algorithm $\mathcal{A}$}
The choice of algorithm $\mathcal{A}$ depends strongly on $\mathcal{P}$ and the kind of QCVV data the algorithm will have access to.  For example, properties that are categorical would require \emph{classification} algorithms, whereas properties that are continuous would require \emph{regression} algorithms. In addition, if the training data can be labeled with known values of the property, then \emph{supervised} learning algorithms would be appropriate; in contrast, if the data was unlabeled \emph{unsupervised} learning algorithms would be necessary.

ML algorithms can be complicated, but at their heart, they consist of parameters $\bs{\theta}$ that index a hypothesis for the relationship between data and the property of interest. That is, a fixed value of $\bs{\theta}$ defines a \emph{particular} map $f_{\bs{\theta}}: \mathcal{D} \rightarrow \mathcal{P}$. During \emph{training}, the algorithm updates these parameters in response to training data to find an element of the hypothesis class (i.e., a map $f_{\bs{\theta}}$) that can predict the property with high accuracy.

\subsubsection{A performance measure $P$ used to evaluate predictions}
Some measure of performance is necessary to evaluate whether a given map $f_{\bs{\theta}}$ is a good one.  It measures the quality of $f_{\bs{\theta}}$'s predictions.  Typically, the performance measure is defined in terms of a \emph{loss function} to quantify the penalty for an incorrect prediction.  For binary or discrete classification, the loss function is usually ``0/1'': a penalty of 1 is applied for an incorrect classification and no penalty is applied for a correct one. For continuous regression problems, the loss function quantifies the distance between the predicted value of $\mathcal{P}$ and its true value. During training, ML algorithms update the parameters $\bs{\theta}$ to minimize the loss. Phrased another way, training is a search over the hypothesis class to find the best prediction map, as quantified by the performance measure.

The performance of a classifier (map $f_{\bs{\theta}}$) can be easily evaluated: use the classifier to classify a feature vector and compare the label it assigns to the true label. Evaluating the performance of an \emph{algorithm} is a slightly harder task. A common way to do so is to use \emph{cross-validation}. The data is split into two parts, a training set and a testing set. The algorithm then learns using data from the training set, and the performance of the classifier it learns is computed the testing set.

\subsubsection{A specification of (or search protocol over) the algorithm's hyperparameters}
Hyperparameters are user-controllable parameters that affect the algorithm's behavior, and the algorithm's performance depends on their values.  In simple cases the hyperparameters can be specified \emph{a priori}.  In other cases, a procedure for varying the hyperparameters to find good values is necessary. Such \emph{hyperparameter turning} can be time consuming, because the algorithm's performance must be evaluated for each hyperparameter setting. However, without hyperparameter tuning, the accuracy of the prediction map learned by the algorithm may be artificially low.

\subsubsection{Labeled training data from which the algorithm can learn}
ML algorithms learn using training data. Consequently, a machine-learned QCVV technique requires a collection of training data $\mathcal{C}$ that the ML algorithm can learn from. In supervised learning, each data set $\mathcal{D}_{j}$ in the collection carries with it a label defined by an instance of the property of interest, $\mathcal{P}_{j}$.

In practice, $\mathcal{C}$ is a \emph{sample} of all possible data sets that could be observed. For this reason, checking whether $\mathcal{C}$ is \emph{representative} of the infinite-sample collection is vital to ensuring the prediction function learned by the ML algorithm will do well when inferring the property of interest on new  data (i.e., how well the rule will \emph{generalize}). Assuming $\mathcal{C}$ is sufficiently representative, cross-validation over $\mathcal{C}$ is a sensible way of estimating how well the prediction map will generalize to new data.
~\\~\\
Note the first two components in this rubric (the property and the experiment design) need to be specified for \emph{any} QCVV technique.  Several of the others could -- at least in principle -- be discovered themselves by other machine learning algorithms:  the experiment design (at least partly, via reinforcement learning \cite{Sutton}), the feature map $\phi$ (via automated feature engineering/synthesis \cite{Kanter2015}), and the choice of the hyperparameters (via automated hyperparameter tuning \cite{Bergstra2013}).

We now demonstrate exactly how to specify these components for the simple but relevant problem we consider.

\section{A machine-learned QCVV technique for distinguishing single-qubit coherent and stochastic noise}
\label{sec:application}

To demonstrate the above rubric, we consider a simple QCVV task, ``Estimate the coherence of noise." Quantifying the coherence of noise is important because the noise's coherence affects the relationship between average and worst-case error rates \cite{Kueng2016a}, and potentially the performance of quantum error-correcting codes \cite{Barnes2017,Darmawan2017,Wallman2016,Guti2016,Sheldon2016a}. An existing QCVV technique for estimating the noise's coherence is unitarity benchmarking \cite{Wallman2015, Feng2016}. Gate set tomography \cite{Blume-Kohout2013a, Greenbaum2015,Blume-Kohout2017a} can also do so. 

Estimating the noise's coherence is a \emph{regression} task, since the property of interest is a continuous number. While there exist ML algorithms for that type of task, here we instead focus on a highly-idealized and highly-specialized variant of this problem. We assume the noise is \emph{either} coherent \emph{or} stochastic (general noise is a combination of coherent, stochastic, and other types of error), which means the task ``estimate the coherence of the noise" becomes ``determine whether the noise is coherent or stochastic."

This is a classification problem, and well-suited to ML algorithms. Phrased another way, if ML algorithms cannot perform the task ``distinguish coherent from stochastic noise", there's little reason to expect  they can perform the more general task ``estimate the coherence of the noise." We emphasize that our interest in this work lies in answering the question ``Can ML algorithms be used to develop a QCVV technique?", and \emph{not} ``What novel QCVV techniques can be learned by ML algorithms?" As noted in Section \ref{sec:intro}, characterizing NISQ processors is challenging for many reasons. Our goal here is to prove the principle, so we simplify by considering a \emph{single-qubit} QIP.

In the remainder of this section, we specify each of the components of the rubric described in the previous section. Readers interested solely in our results may proceed to Section \ref{sec:clf_gst}.

\subsubsection{Property $\mathcal{P}$: ``Are the gate errors coherent or stochastic?"}
\label{sec:qipnoise}

The single-qubit QIP we consider will be assumed to have five operations:  initialization of some fiducial state $\rho_{0}$,  three logic gates corresponding to idling and $\pi/2$ rotations around the $X$ and $Y$ axes of the Bloch sphere ($G_I, G_X$, and $G_Y$, respectively), and a terminating measurement, $E$. The collection $\{\rho_{0}, E, \{G_I, G_X, G_Y\}\}$ we refer to as the \emph{gate set} of the QIP.

An ideal unitary gate $G_0$ is described by a \emph{completely positive, trace-preserving} (CPTP) map. A common representation of this action is $G_0[\rho] = U\rho U^\dagger$, where $U$ is \emph{generated} by a $2\times 2$ Hermitian matrix $H_{0}$: $U = e^{-iH_{0}}$. A different representation of that same unitary action is $G_{0}[\rho] = e^{\mathcal{H}_{0}}[\rho]$, where now $\mathcal{H}_0$ is a $4\times 4$ matrix acting on density matrices as $\mathcal{H}_0[\rho] = -i[H_0,\rho]$.

The real gate $\mathcal{E}$ will deviate from $G_0$.  This deviation is the \emph{error} in the gate, and a variety of errors are possible.  As noted above, we are interested in two specific classes of errors, and consider an idealized scenario where the gate errors are either fully coherent, or fully stochastic. To make this precise, we define the following models of fully coherent and fully stochastic gate errors, which we describe by their generators.

If $\mathcal{E}$'s error is purely \emph{coherent}, then it still acts unitarily, but with an additional Hamiltonian term, $H_{e}$:
\begin{equation}
G_{0}[\rho] \rightarrow \mathcal{E}[\rho] \equiv e^{-i(H_{0}+H_{e})}\rho e^{i(H_{0}+H_{e})}.
\end{equation}
This action can be also written as 
\begin{equation}
\mathcal{E} = e^{\mathcal{H}_0 + \mathcal{H}_e},~\text{where}~\mathcal{H}_{e}[\rho] = -i[H_{e}, \rho].
\end{equation}
An error is purely coherent if and only if it is of this form.  Coherent errors often result from imperfect calibration of classical control fields.

Regarding ``stochastic'' noise, we use a particular definition generally consistent with the way the word is used in the quantum computing community.  The concept our definition captures is this:  stochastic errors are what occur when control fields are \emph{fluctuating} around the desired value in a random way, but the \emph{expected value} of those fluctuations is zero.

This concept is most straightforwardly seen when considering an imperfect idle operation (where the target gate is $G_{I}$). This operation has purely stochastic errors if its noisy version can be written as a convex combination of unitary operations \emph{and} it is invariant under time reversal:
\begin{align}
\mathcal{E}[\rho] &= \sum_{j=1}^{n}w_{j}U_{j}\rho U_{j}^{\dagger},\\
\mathcal{E} &= \mathcal{E}^\dagger.
\end{align}
We generalize this definition to nontrivial unitary gates $G_{0} = e^{\mathcal{H}_{0}}$ via the generators for stochastic noise appearing in the canonical Lindblad equation.  Letting $\mathcal{S}$ represent a parameterized generator, then the noisy gate is
\begin{equation}
\mathcal{E} = e^{\mathcal{H}_0 + \mathcal{S}}.
\end{equation}
Details of these definitions are given in Appendix \ref{app:noisesim}.  Stochastic errors can have several causes, including fluctuations of the classical control fields and weak coupling to a rapidly mixing quantum bath.

\begin{figure}
\includegraphics[width=\columnwidth]{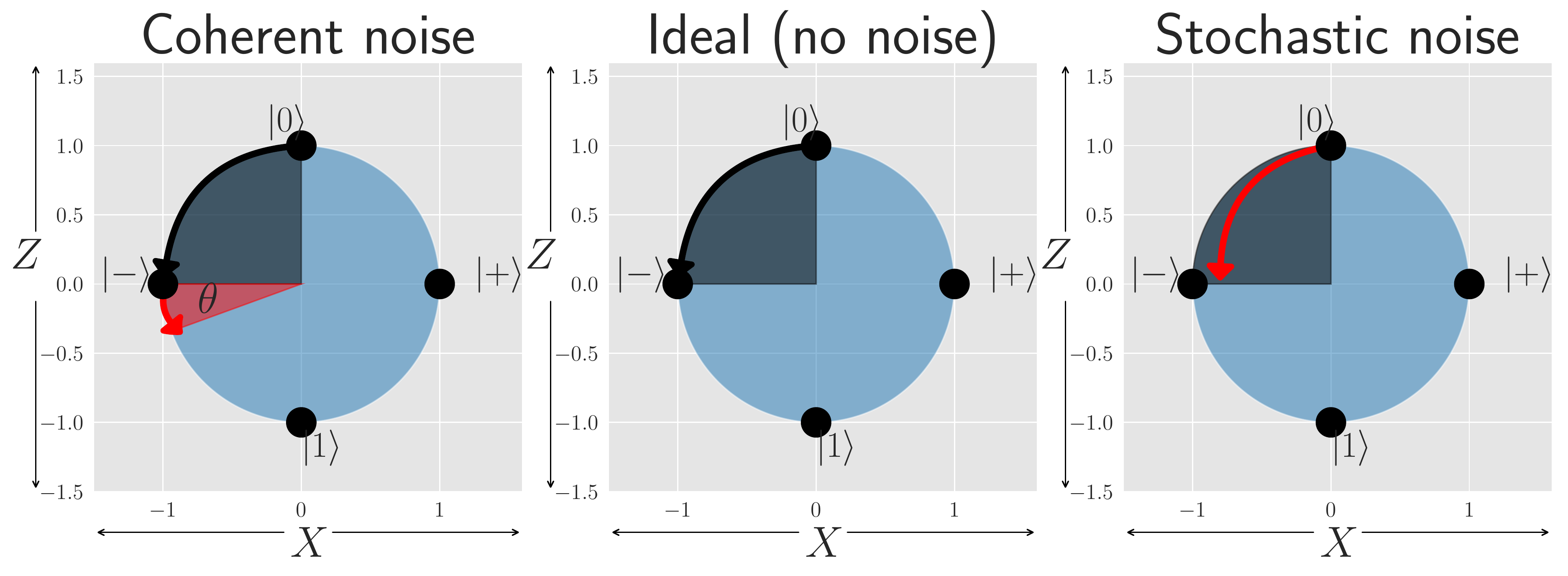}
 \caption{\textbf{Example: realizations of coherent and stochastic noise.} Suppose $\rho_{0} = |0\rangle\langle 0|$, $G_{0} = R_{Y}(\alpha)$ is a rotation about the $Y$-axis by an angle $\alpha = \pi/2$, and the POVM is $\{|0\rangle\langle 0| , |1\rangle\langle 1|\}$. The pictures above show a cross-section of the qubit Bloch sphere under the action of: the ideal gate (center), a gate with a coherent over-rotation by an angle $\theta$ (left), and a gate with a simple stochastic noise model (right).}
\label{fig:noise3}
\end{figure}

Both of these noise models are \emph{Markovian}: for a fixed realization of the noise, wherever $G_{0}$ occurs in a circuit the QIP is supposed to perform, the \emph{same} CPTP map $\mathcal{E}$ is applied by the QIP. Figure \ref{fig:noise3} shows specific examples (\emph{realizations}) of how purely coherent or purely stochastic noise affect a simple circuit for a single qubit.

As discussed in Section \ref{subsec:data_set_descrip}, we generate random realizations of stochastic and coherence noise for training and testing. Details of our simulation method are given in  Appendix \ref{app:noisesim}. In generating realizations of these two noises we control the \emph{strength} of the error with a parameter $\eta$.  The coefficient or ``rate'' $r$ that multiplies each generator of stochastic or coherent noise is randomly distributed, and $\expect{|r|} = O(\eta)$.  For small $\eta$, this means $|\mathcal{E} - G_{0}| = O(\eta)$ as well.

\subsubsection{Experiment design: gate set tomography (GST) circuits}
\label{sec:gst_overview}

We use a general purpose experiment design that provides information about any Markovian property; namely, the circuits prescribed for gate set tomography (GST) \cite{Blume-Kohout2013a, Greenbaum2015,Blume-Kohout2017a}.   This ensures the property we are interested in can be inferred from experimental data. GST seeks to completely reconstruct the QIP's gate set, and so its experiment design must be sensitive to everything about Markovian noise -- including the binary ``coherent vs. stochastic" property. The data we consider are the outcome probabilities (or, in the case of real finite-sample data, the observed \emph{frequencies}) of the circuits used for GST. Each circuit $c_{j}$ in the GST experiment design is of the form $F_{M}\circ(g_{m})^{l}\circ F_{S}$ where $F_{S},F_{M}$, and $g_{m}$ are short subcircuits comprised of the elementary gates from the QIP's gate set \footnote{We use ``$\circ$" to denote composition of the channels, and use the notation $(g_{m})^{l}$ to mean $g_{m}$ composed with itself $l$ times.}, and the ``germ" $g_{m}$ is repeated $l$ times.

GST doesn't prescribe a single, specific experiment design, because it has some configurable parameters.  One is the \emph{maximum depth} of the longest circuit in the experiment design, denoted $L$. The experiment design includes circuits with $l=1,2,4,...$ repetitions of each germ, but only up to the point where the depth of $g_{m}^{l}$ exceeds $L$. Increasing $L$ adds new, longer circuits to the experiment design, which amplify noise more.

For a fixed value of $L$, the circuits in the experiment design have depth at most $L + O(1)$. In what follows, we use $L$ as an index to specify the experiment design. For instance, the ``$L=1$ experiment design" consists of all the circuits used in $L=1$ GST, and so on for higher values of $L$.

Each circuit $c_j$ terminates with a measurement $M$. For single-qubit GST, $M$ is a 2-outcome POVM, with outcomes ``0" and ``1". The random outcomes of running each experiment in the experiment design constitute the experimental data. When analyzing this data, we often assume the noise is stationary, meaning the \emph{order} of the outcomes is irrelevant. This means all the information about the noise is contained in the circuit's outcome probability $p_{j}$, which can be estimated by counting the number of times a given outcome was observed, and dividing by the total number of times the circuit was repeated, $N_{\mathrm{samples}}$:
\begin{equation}
\hat{p}_{j} = f_{j} \equiv \frac{\text{\# ``0'' outcomes seen when running $c_{j}$}}{N_{\mathrm{samples}}}.
\end{equation}

For a given experiment design, a GST data set $\mathcal{D}$ is a list $[N_{\mathrm{samples}}, (c_{1}, f_{1}), (c_{2}, f_{2}), \cdots, (c_{d}, f_{d})]$. Here, $d$ is the total number of circuits (\emph{not} the Hilbert space dimension, which is 2 throughout this paper!), which depends on the index $L$.

\subsubsection{Feature space $\mathcal{F}$: the unit hypercube}
\label{subsec:gst_feature_space}

A GST data set is usually presented as a list of count statistics, one for each circuit.  But ML algorithms represent data as \emph{feature vectors}.  So we need to map a GST data set into a \emph{feature vector} $\mathbf{f}$ in a \emph{feature space} $\mathcal{F}$.  Any way of doing so is called a \emph{feature map}, denoted $\phi$.

The most obvious feature map comes from arranging the estimated outcome probabilities for each circuit into a vector:
\begin{align}
\label{eq:feature_map_1}
\phi(\mathcal{D}) \equiv \mathbf{f} &=  (f_{1}, f_{2}, \cdots, f_{d})\\
\phi &: \mathcal{D} \rightarrow \mathbf{f} \in [0,1]^{d}.
\end{align}
The feature space defined by this $\phi$ is the unit hypercube in $\mathbb{R}^{d}$:
\begin{equation}
\mathcal{F} =  [0, 1]^{d}.
\end{equation}

\begin{figure}
\includegraphics[width=\columnwidth]{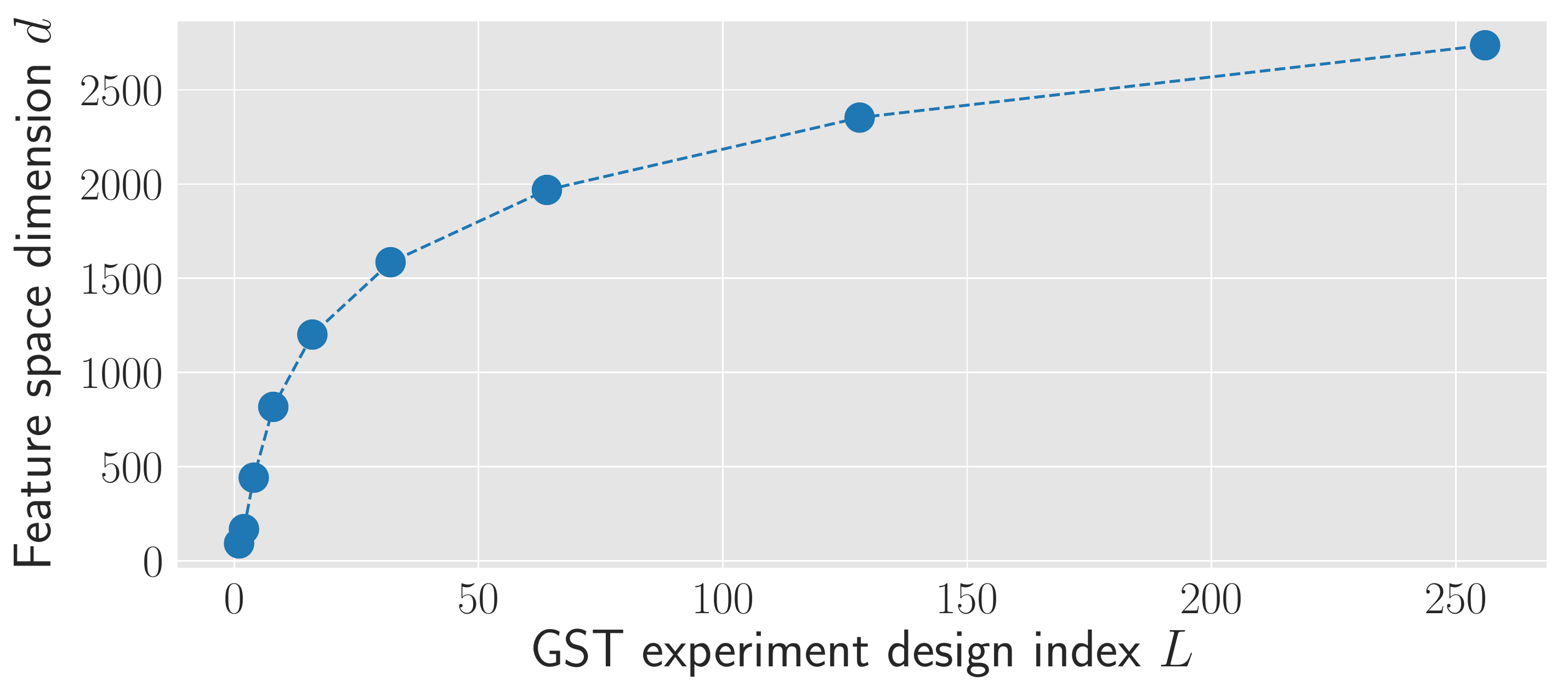}
 \caption{\textbf{Feature space dimension $d$ grows with the GST experiment design index $L$}. We use a simple feature map $\phi$ to embed GST data sets into a ``base" feature space $\mathcal{F}$. As the experiment design index $L$ is increased, the number of circuits in the experiment design grows, which increases the dimension of the base feature space. This growth is roughly logarithmic in $L$, and even at the largest value of $L$ considered here, the feature space dimension is not too large for the ML algorithms we use.}
\label{fig:feature_vec_dim}
\end{figure}

The dimension of $\mathcal{F}$ is $d$, the total number of circuits in the experiment design. This varies with $L$.  If $\mathcal{D}$ is generated from an $L=l$ experiment design, we call $\mathbf{f}$ an ``$L=l$ GST feature vector". Figure \ref{fig:feature_vec_dim} shows $d$ versus $L$, which shows that $d$ grows logarithmically with $L$. Moreover, even the largest GST experiment designs considered here are quite manageable for ML algorithms, which can learn well on feature spaces that have dimension up to $d=10^{5}$.

\begin{figure}
\includegraphics[width=\columnwidth]{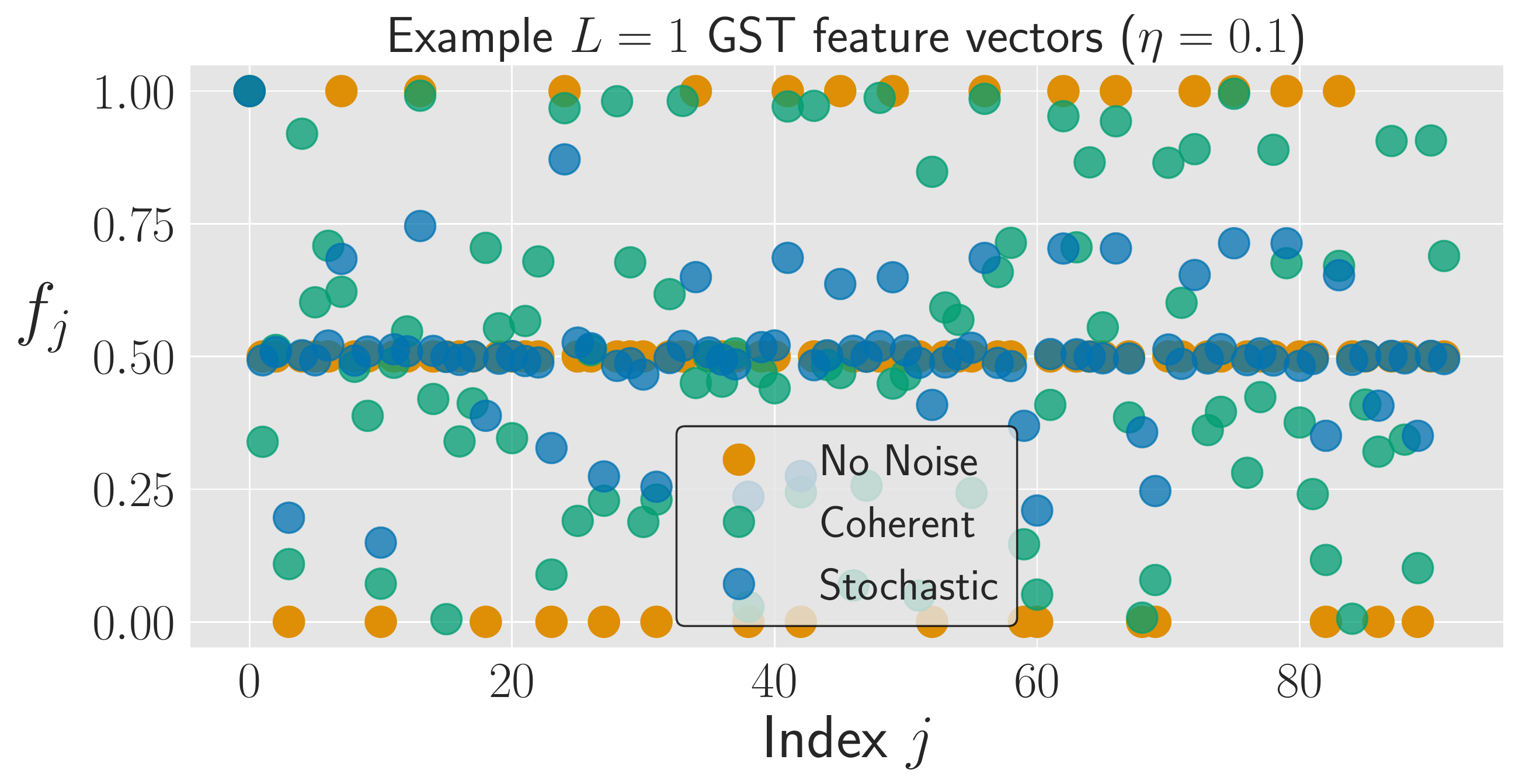}
 \caption{\textbf{GST data sets as feature vectors}. For the $L=1$ GST experiment design, we plot the (infinite-sample/exact) feature vectors for: a noiseless gate set (orange), a gate set where each gate has been affected by an independent coherent error (green), and a gate set where each gate has been affected by an independent stochastic error (blue). Both realizations of the noise strength $\eta = 0.1$.}
\label{fig:noise2}
\end{figure}

Figure \ref{fig:noise2} plots examples of $L=1$ GST feature vectors, and shows how the components of the feature vector change under particular realizations of coherent and stochastic noise.

The ``base'' feature space defined above is just a starting point.  In principle, it contains all the necessary information for classifying whether the noise is coherent or stochastic (since GST circuits are designed to capture and amplify every property of Markovian noise).  As we shall see, that information may not be \emph{easily accessible} to linear classifiers, meaning new feature maps will be defined to enable those algorithms to learn using GST feature vectors.

\subsubsection{Algorithm $\mathcal{A}$: supervised binary classifiers}
\label{sec:binary_clfs}
We use \emph{supervised} classifiers because synthetic training data can be easily generated. The general task of supervised learning (for binary classification) is: ``Given a collection of feature vectors $\mathcal{C} = \{(\mathbf{f}_{j}, y_{j})\}_{j=1}^{N}$, with $y_{j} \in \{\pm 1\}$ indicating which class  the feature vector $\mathbf{f}_{j}\in \mathcal{F}$ belongs to, learn a classifier $c: \mathcal{F} \rightarrow \{\pm 1\}$."

A classifier is a decision surface: the label assigned by the classifier depends on which side of the decision surface the feature vector falls on. All such algorithms seek a classifier that performs well; they differ in (1) how they search for a good decision surface, and (2) how they try to avoid overfitting.

Every binary classification algorithm learns a decision surface dividing the feature space into two parts. \emph{Linear} classifiers are particularly simple: the decision surface is an affine hyperplane. It can be described by a normal vector and a scalar offset.  We focus primarily on linear classifiers here. We also consider two intrinsically nonlinear classifiers.  Because linear classifiers are so important to our analysis, we now briefly review the geometry of affine hyperplanes.
~\\
\paragraph{Affine hyperplane geometry}
~\\

Consider a fixed vector $\bs{\beta}\in \mathbb{R}^{d}$ and a fixed scalar $\beta_{0}$. Together, these define an affine hyperplane $H$; namely, the set of vectors $\mathbf{x}_{j} \in \mathbb{R}^{d}$ satisfying $\boldsymbol{\beta} \cdot \mathbf{x}_{j} + \beta_{0} = 0$. The distance from any point $\mathbf{x}\in \mathbb{R}^{d}$ to any point $\mathbf{x}_{0} \in H$ is
\begin{equation}
 d(\mathbf{x}, \mathbf{x}_{0}) = |\boldsymbol{\beta}\cdot\mathbf{x} +\beta_{0}|/||\boldsymbol{\beta}||.
\end{equation}
As the relation above is independent of $\mathbf{x}_{0}$, we may refer to ``the distance from $\mathbf{x}$ to $H$" without ambiguity, which we denote as $d(\mathbf{x}, H)$.

The classification rule $c$ learned by a linear classifier is an affine hyperplane:
\begin{equation}
\label{eq:linear_clf}
c(\mathbf{f}) = \text{sign}\left[\boldsymbol{\beta}\cdot \mathbf{f} + \beta_{0}\right].
\end{equation}

Suppose $H$ separates a collection of labeled feature vectors $\mathcal{C}$ (i.e., $c(\mathbf{f}_{j}) = y_{j}~\forall~j$). The \emph{geometric margin} of $H$ (hereafter, ``margin of $H$"), $M_{H}$, is the minimum distance from any feature vector to $H$:
\begin{equation}
\label{eq:geom_margin}
M_{H} = \underset{\mathbf{f}_{j} \in \mathcal{C}}{\min}~d(\mathbf{f}_{j}, H) = \frac{1}{||\boldsymbol{\beta}||}\underset{\mathbf{f}_{j} \in \mathcal{C}}{\min}~\left|\boldsymbol{\beta}\cdot \mathbf{f}_{j} + \beta_{0}\right|.
\end{equation}
Suppose $H$ is a fixed separating hyperplane for $\mathcal{C}$. Its margin $M_{H}$ has a nice geometric interpretation: any feature vector perturbed by an amount greater than $M_{H}$ along $\hat{\bs{\beta}}$ would be misclassified by $H$. As such, $M_{H}$ relates to the \emph{robustness} of the decision surface learned by the classifier: higher-margin hyperplanes are more robust to perturbations of the data. Hence, a large-margin hyperplane is desirable.  When there are multiple hyperplanes that separate the training data, the \emph{optimal} hyperplane is the one that maximizes $M_{H}$. If $H$ has the largest geometric margin of all hyperplanes that could separate $\mathcal{C}$, then it follows that $c$ would generalize well when classifying feature vectors that are small perturbations on the feature vectors in $\mathcal{C}$.  The margin becomes especially important when the training data are limited, or when the QCVV data to be classified has finite sample noise (see Section \ref{sec:finite_sampling}).

~\\
\paragraph{Algorithms for supervised binary classification}
\label{subsub:ml_algs}
~\\

We examine and compare the performance of several algorithms for supervised binary classification. We consider five simple, widely-used algorithms.  Three are linear classifiers:  linear discriminant analysis (LDA), perceptrons, and linear support vector machines (SVMs).  Two are intrinsically nonlinear: quadratic discriminant analysis (QDA), and radial basis function (RBF) SVMs.  We explain each of these algorithms below.  Further discussion can be found in Hastie \emph{et al.} \cite{Hastie2008}. We used the implementations of these algorithms available in the open-source Python package scikit-learn \cite{Pedregosa2011}.

One class we do \emph{not} consider is neural networks.  Neural networks have a complex internal structure that makes them very powerful, but their behavior is difficult to understand and explain.  

\emph{LDA} and \emph{QDA} approach binary classification from a statistical perspective.  They are derived from a Gaussian ansatz that assumes the feature vectors for each class are normally-distributed with means $\bs{\mu}_{1}$ and $\bs{\mu}_{2}$ and covariances $\Sigma_{1}$ and $\Sigma_{2}$. Under this assumption, the optimal decision surface can be derived from a likelihood ratio test.  ``Training'' these algorithms simply means assuming that the means and covariances are unknown, and estimating them from the training data. We denote the estimated means and covariances by $\hat{\bs{\mu}}_{j}$ and $\hat{\Sigma}_{j}$, respectively.

LDA assumes the two covariance matrices are identical, in which case the optimal decision surface is a hyperplane. The classification rule learned by the LDA algorithm is
\begin{align}
\begin{split}
c_{\mathrm{LDA}}(\mathbf{f}) &= \mathrm{sign}\left[\mathbf{f}^{T}\hat{\Sigma}^{-1}(  \hat{\boldsymbol{\mu}}_{1} -  \hat{\boldsymbol{\mu}}_{2})\right.\\
&\left.~~~~~~~~~~+  (\hat{\boldsymbol{\mu}}_{1}^{T}\hat{\Sigma}^{-1} \hat{\boldsymbol{\mu}}_{1} -\hat{\boldsymbol{\mu}}_{2}^{T}\hat{\Sigma}^{-1} \hat{\boldsymbol{\mu}}_{2})/2 \right].
\end{split}
\end{align}
QDA allows the two covariance matrices to be different, which produces an optimal classification rule of the form
\begin{align}
\nonumber c_{\mathrm{QDA}}(\mathbf{f}) &= \text{sign}\left[\log\left(\frac{|\hat{\Sigma}_{2}|}{|\hat{\Sigma}_{1}|}\right) + (\mathbf{f} - \hat{\boldsymbol{\mu}}_{2})^{T}\hat{\Sigma}_{2}^{-1}(\mathbf{f} - \hat{\boldsymbol{\mu}}_{2})\right.\\
&\left.~~~~~~~~~~- (\mathbf{f} - \hat{\boldsymbol{\mu}}_{1})^{T}\hat{\Sigma}_{1}^{-1}(\mathbf{f} - \hat{\boldsymbol{\mu}}_{1})\right].
\end{align}
The classification rule for QDA is equivalent to LDA if $\hat{\Sigma}_{1} = \hat{\Sigma}_{2} $.

Estimating the means $\bs{\mu}_{1}$ and $\bs{\mu}_{2}$ is relatively straightforward as long as there are significantly more training examples than the feature space dimension $d$.  But unbiased estimation of the covariance matrices would require much larger training sets (more than $d^2$ examples), so both LDA and QDA \emph{regularize} the estimated covariance matrices. LDA does so through a hyperparameter $\tau$ affecting the dimension of $\bs{\beta}$, by controlling the rank of $(\hat{\Sigma})^{-1}$. QDA introduces a hyperparameter $s$, where $0\leq s \leq1$, and replaces $\hat{\Sigma}_{k}$ by  $s\hat{\Sigma}_{k} + (1-s)\mathcal{I}$.

The \emph{perceptron} \cite{Rosenblatt1958, Rosenblatt1957} is one of the oldest supervised binary classification algorithms. It makes no assumptions about the distribution of the feature vectors, but learns a separating hyperplane using a simple, iterative training algorithm whose only hyperparameter, $N_{\mathrm{epochs}}$, determines the maximum number of iterations before the algorithm terminates.  The result is a linear classification rule:
\begin{equation}
c_{\text{Perceptron}}(\mathbf{f}) = \mathrm{sign}\left[\boldsymbol{\beta}\cdot \mathbf{f} + \beta_{0}\right].
\end{equation}

\emph{Support vector machines} (SVMs) \cite{Vapnik1964} address a notable flaw of the perceptron algorithm: the perceptron will find \emph{some} separating hyperplane (if one exists), but not necessarily an \emph{optimal} (``maximal-margin") one. The \emph{soft-margin} SVM explicitly optimizes the margin of the hyperplane it learns, subject to a regularization penalty $C$. The classification rule learned by the linear SVM is
\begin{equation}
c_{\text{Linear SVM}}(\mathbf{f}) = \text{sign}\left[\sum_{j=1}^{N}y_{j}c_{j}(\mathbf{f}_{j}\cdot \mathbf{f}) + \beta_{0}\right],
\end{equation}
where $0 \leq c_{j} \leq C$. The regularization penalty $C$ is a hyperparameter for this algorithm, and controls a trade-off between maximizing $M_{H}$ and minimizing the number of mis-classified points. The classification rule explicitly depends on the some of the feature vectors in $\mathcal{C}$: if $c_{j} \neq 0$, the corresponding feature vector $\mathbf{f}_{j}$ is called a \emph{support vector}.

The \emph{radial basis function (RBF) SVM} is intended for situations where the data can only be separated by highly-curved (and thus, highly-nonlinear) hypersurfaces.  The RBF SVM uses a kernel $K: \mathcal{F}\otimes \mathcal{F} \rightarrow \mathbb{R}$ \cite{Hofmann2008, Guyon1993} to implicitly map the data to a high-dimensional feature space where a linear decision surface can separate the data. In the original feautre space, the surface is \emph{nonlinear}. The RBF SVM thus implicitly performs a particular kind of feature engineering, akin to what we do explicitly later in this paper (Section \ref{subsec:feature_engineering}).  The RBF SVM classification rule is
\begin{equation}
\label{eq:rbfsvm}
c_{\text{RBF SVM}}(\mathbf{f}) = \text{sign}\left[\sum_{j=1}^{N}y_{j}c_{j}K(\mathbf{f}_{j}, \mathbf{f}) + \beta_{0}\right],
\end{equation}
where $K(\mathbf{x}, \mathbf{y}) = \text{exp}\left[-\gamma || \mathbf{x} - \mathbf{y}||^{2}\right]$, and again $0 \leq c_{j} \leq C$. The RBF SVM has two hyperparameters: $C$ (which plays the same role as in the linear SVM), and $\gamma$, which controls the \emph{width} of the kernel.

\begin{table*}
\begin{center}
  \begin{tabular}{|p{6cm}|p{8cm}|}
  \hline
  Target gate set $\mathcal{G}$& $\rho_{0} = |0\rangle\langle 0|$, $\{G_{I}, G_{X}, G_{Y}\}$, $E = \{|0\rangle\langle 0|, |1\rangle\langle 1|\}$\\\hline
   Noise type & Coherent or stochastic (see Appendix \ref{app:noisesim})\\\hline
   GST experiment design index $L$ & $(1,   2,   4,   8,  16,  32,  64, 128, 256)$\\\hline
   Noise strength $\eta$ &$\{1, 2.15,   4.64\} \times \{10^{-4}, 10^{-3}, 10^{-2}\}, \newline \{1, 1.19, 1.43, 1.71, 2.04, 2.44, 2.92, 3.49, 4.18, 5\} \times 10^{-1}$ \\\hline
   Noise realizations for fixed $\eta$ and noise type & 900 (1 for each gate in $\mathcal{G}$, yielding 300 noisy gate sets, generated as specified in Appendix \ref{app:noisesim})\\\hline
  \end{tabular}
  \caption{\textbf{Data set description.}
  For each value of $L$, a collection of data $\mathcal{C}_{L}$ was generated using noisy versions of the target gate set $\mathcal{G}$. The noise strength $\eta$ quantifies the discrepancy between the ideal gate set and its noisy version.}
    \label{tab:data_descrip}
    \end{center}
 \end{table*}

\subsubsection{Performance measure $P$}
To search over its hypothesis class, an ML algorithm needs a measure of how good any hypothesis in the class is. Here, we use a simple performance measure for a given hypothesis:
\begin{equation}
\label{eq:accuracy}
A= \begin{cases}1~~\text{if}~~ c(\mathbf{f}_{j}) = y_{j}\cr 0~~\text{otherwise}\end{cases},
\end{equation}
for which the \emph{average} accuracy equals the probability of correct classification. The quantity $(1-A)$ is the ``0/1 loss" introduced in Section \ref{sec:ml_qcvv_framework}.

We typically use a ``$K$-fold shuffle-split" cross-validation approach when evaluating an algorithm: the data is split $K$ times into training and testing data sets (with resampling). We usually take the number of feature vectors in the testing set to be 10\% of the feature vectors in the data collection.
 
\subsubsection{Hyperparameter specification}

\begin{table*}
\begin{center}
  \begin{tabular}{ | l | c | c | c |}
    \hline
  \textbf{Classification algorithm} & \textbf{Hyperparameter}& \textbf{Default value} & \textbf{Values used in this work} \\\hline
    LDA & $\tau$ & $10^{-4}$ & $10^{-5}, 10^{-4}, 10^{-3},10^{-2}$, .1, .25, .5, .75, 1\\ \hline
    QDA & $s$ & 0 &  0, .25, .5, .75, 1\\\hline
    Linear SVM & $C$ &1 & 1,2,5,10,20,50,75,100,150,200,250\\
    \hline
        RBF SVM & $C$ &1&1,2,5,10,20,50,75,100 \\
    \hline
    &$\gamma$ & $1/d$& .01, .1, 1, 10, 100\\\hline
      Perceptron & $N_{\mathrm{epochs}}$  & 5 & 5, 50, 100 250, 300, 500, 750, 1000 \\
    \hline
  \end{tabular}
  \caption{\textbf{Algorithm hyperparameters.} To do hyperparameter tuning, we use a brute-force grid search over the hyperparameter values above. We choose the hyperparameter values that give the best performance. (Default values are those used in scikit-learn 0.19.1.)}
    \label{tab:hyperparams}
       \end{center}
 \end{table*}
 
All of the algorithms discussed in Section \ref{subsub:ml_algs} have hyperparameters that affect their behavior.
For each algorithm, we performed hyperparameter tuning using brute-force grid search over the values listed in Table \ref{tab:hyperparams}; see Appendix \ref{app:hyperparamsweep} for details.

\subsubsection{Data collection $\mathcal{C}$}
\label{subsec:data_set_descrip}

We generated a large collection of labeled training data -- simulated GST datasets for many realizations of purely stochastic and coherent noise -- which we use for training and cross-validation. To produce this data, we numerically simulated GST data sets for several different values of $L$, using noisy gate sets. All of the noisy gate sets were perturbations around a noiseless target gate set $\mathcal{G} = \{\rho_{0} = |0\rangle\langle 0|, \{\mathcal{I}, X_{\pi/2}, Y_{\pi/2}\}, E = \{|0\rangle\langle 0|, |1\rangle\langle 1\}\}$.

We chose 19 values for the noise strength $\eta$ (see Table \ref{tab:data_descrip}). For each noise type (stochastic or coherent) and noise strength $\eta$, we generated 300 random noisy gatesets (see Appendix \ref{app:noisesim}), each one obtained by adding a randomly-sampled error generator to the generator for each gate in the gate set.  We used these noisy gate sets to generate GST data sets using pyGSTi \cite{Nielsen2017}. Our main focus is on the $N\to\infty$ exact-sampling limit, where the estimated frequencies equal the exact outcome probabilities, but we consider finite-sample effects in Section \ref{sec:finite_sampling}.

The collection of labeled feature vectors generated this way, for a specific value of $L$, is denoted $\mathcal{C}_L$.  So $\mathcal{C}_{L} = \{(\mathbf{f}_{j}, y_{j})\}_{j=1}^{N}$, where $\mathbf{f}_{j} = \phi(\mathcal{D}_{j}) \in \mathbb{R}^{d}$, $d$ depends on $L$, and $y_j$ is the binary label indicating ``stochastic'' or ``coherent''.  For each value of $L$, there are $N = 2 \times 19\times 300 = 11400$ labeled feature vectors that can be used for training or testing.

Many ML algorithms perform best on data with zero mean and unit variance. So we \emph{standardized} the feature vectors in $\mathcal{C}_{L}$ so that they had this property. The standardized feature vectors are easily computed: let $\hat{\bs{\mu}}$ ($\hat{\Sigma}$) denote the estimated mean (covariance) of the feature vectors. For each $\mathbf{f}_{j}$, its standardized version is
\begin{equation}
\mathbf{f}_{j} \rightarrow  \mathrm{diag}(\hat{\Sigma})^{-1}(\mathbf{f}_{j} - \hat{\bs{\mu}}).
\end{equation}
Using this standardization technique, $\langle \mathbf{f} \rangle = \mathbf{0}$ and $\text{Cov}(\mathbf{f}_{j},\mathbf{f}_{k}) = \mathcal{I}$. We scale by $\mathrm{diag}(\hat{\Sigma})$ rather than $\hat{\Sigma}$ itself so as to adjust each component of $\mathbf{f}$ independently, meaning no correlation is introduced between the features.

\section{Results}
\label{sec:clf_gst}

Our goal was to determine how ML algorithms can be usefully applied to QCVV tasks. In this section, we present results showing supervised classifiers can, in fact, learn a high-accuracy decision surface for distinguishing between coherent and stochastic noise, but that their accuracy depends strongly on both the GST experiment design \emph{and} the use of feature engineering.

\subsection{Classifying GST feature vectors}
\subsubsection{Testing whether linear classification is feasible}
\label{subsec:feasibility_testing}

Three of the classification algorithms presented in Section \ref{sec:binary_clfs} learn a linear decision surface. If $\mathcal{C}_{L}$ is not linearly separable, then those algorithms cannot perform well. So for each $L$, we began our analysis by determining whether $\mathcal{C}_{L}$ was linearly separable. In the literature, this is usually checked by running the perceptron algorithm with $N_{\mathrm{epochs}} >> 1$. If the algorithm converges to a separating hyperplane, then clearly $\mathcal{C}_{L}$ is linearly separable. However, a failure to converge does not guarantee that $\mathcal{C}_{L}$ is linearly \emph{inseparable}.  

Instead of using the perceptron to check for linear separability, we constructed a linear program to test for separability that either finds a separating hyperplane \emph{or} (if the data are inseparable) constructs a provable witness to that fact (Appendix \ref{app:lp_problem}).

We found that each $\mathcal{C}_L$ was linearly separable \emph{except} for $L=1$ (``linear GST'' data).  Therefore, no linear classification algorithm can attain 100\% accuracy in classifying coherent and stochastic noise using $L=1$ GST feature vectors, even in principle.  This says something nontrivial about the \emph{geometry} of the hypersurfaces that could separate the data: they aren't hyperplanes (see Section \ref{subsec:dim_reduction}).

However, this only holds true when the range of values for the noise strength $\eta$ is quite large, extending  3 orders of magnitude from $10^{-4}$ to $0.5$.  We found that subsets of $\mathcal{C}_{1}$ with a \emph{restricted} range of $\eta$ were almost always linearly separable (see Table \ref{tab:lp_results_L_1}).  We conclude that this QCVV problem gets harder (at least somewhat) when the gate errors are allowed to be very large.  This is unsurprising; similar challenges afflict randomized benchmarking and GST, and most QCVV methods focus on the regime where gate errors are perturbative.

For $L>1$, $\mathcal{C}_{L}$ was always linearly separable.  However, we considered the possibility this might have been an artifact of finite data (undersampling of the noise realizations), rather than an indication that the two noise classes can always be separated using linear classifiers. That is, $\mathcal{C}_{L}$ may not have been a representative sample of all possible feature vectors.

To check for this, we trained a linear soft-margin SVM ($C=10^{4}$) using the full $11400$ feature vectors in $\mathcal{C}_{L}$, and then tested the accuracy of the decision surface it learned on $20900$ previously-unseen feature vectors. The accuracy remains stable between the training and testing data (see Table \ref{tab:undersampling_results}), suggesting that for $L>1$, $\mathcal{C}_{L}$ is in fact linearly separable, and that we were not undersampling the noise realizations. Note that in the following analysis, we do not use the extra $20900$  feature vectors.

\begin{table}
\begin{center}
  \begin{tabular}{|c|c|}
    \hline
Value(s) of $\eta$ in subset & Subset is separable? \\\hline
$[10^{-4}, 0.5]$  (full range)& \textbf{No}\\\hline
Fixed & Yes \\\hline
$[10^{-4}, 0.34]$ &  Yes \\\hline
$[10^{-4}, 10^{-3}]$ & Yes  \\\hline
$[10^{-4}, 10^{-2}]$ & Yes  \\\hline
$[10^{-2}, 10^{-1}]$ & Yes  \\\hline
  \end{tabular}
  \caption{\textbf{Testing for linear separability of subsets of $\mathcal{C}_{1}$}. Over the full range of $\eta$ ($[10^{-4}, 0.5]$), the data collection $\mathcal{C}_{1}$ is linearly \emph{inseparable}. However, subsets of $\mathcal{C}_{1}$ for a \emph{fixed} value of $\eta$, or where $\eta$ varies over a restricted range,  are linearly separable.}
 \label{tab:lp_results_L_1}
 \end{center}
 \end{table}

\begin{table}
\begin{center}
  \begin{tabular}{|c|c|c|}
    \hline
L & Train accuracy & Test accuracy\\\hline
2 & 0.9997 &0.976\\\hline  
4 & 1.0 & 0.963 \\\hline
8 & 1.0 & 0.965 \\\hline
16 & 1.0 & 0.975 \\\hline
\end{tabular}
  \caption{\textbf{Determining whether noise realizations were undersampled}. To evaluate whether we had undersampled the noise realizations -- and were thereby erroneously concluding $\mathcal{C}_{L}$ was linearly separable for $L>1$ -- we trained a soft-margin SVM on all feature vectors in $\mathcal{C}_{L}$ and tested the accuracy of the hyperplane it learned on $20900$ \emph{previously-unseen} feature vectors. The stability between the test and train accuracies suggests we have not undersampled the noise realizations.}
 \label{tab:undersampling_results}
 \end{center}
 \end{table}
 
\subsubsection{Classification accuracy depends on $L$, and hyperparameter tuning is necessary}
\label{subsection:clf_results}

\begin{figure}
\includegraphics[width=\columnwidth]{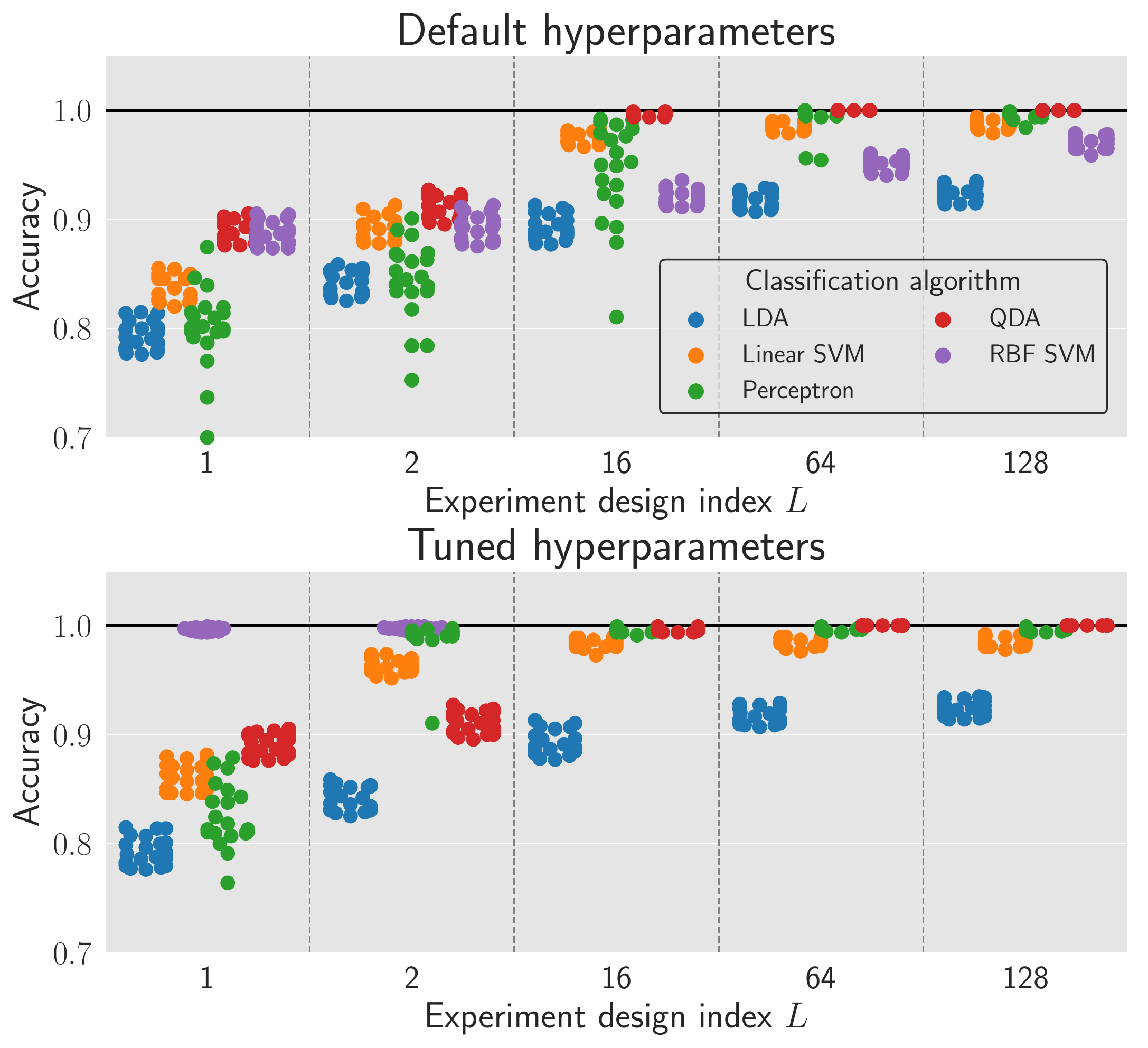}
 \caption{\textbf{Swarmplot of $K=20$ cross-validated classification accuracies as a function of $L$.} As $L$ increases, the accuracy increases in a classifier-dependent way. \textbf{Top}: Under default hyperparameters, the QDA algorithm typically performs best. \textbf{Bottom}:  Hyperparameter tuning boosts the accuracy of the linear SVM, perceptron, and RBF SVM algorithms.}
\label{fig:cv_1}
\end{figure}

When $\mathcal{C}_{L}$ is linearly separable, a linear classifier could in principle successfully learn a separating hyperplane. Whether such an algorithm succeeds in \emph{practice} depends very much on its \emph{hyperparameters}. Figure \ref{fig:cv_1} plots the cross-validated accuracies of the ML algorithms as a function of the GST experiment design index $L$. In the top panel, no hyperparameter tuning was performed, while in the bottom, grid search was used to identify good hyperparameter choices. (See Appendix \ref{app:hyperparamsweep} for results of the grid search.)

In general, classification accuracy increases with $L$. This should be expected, because as $L$ goes up, 
additional circuits are added to the experiment design whose outcome probabilities depend on the underlying gates in an increasingly non-linear fashion. These extra features give the classifier a richer set of data to learn from.

For $L \geq 16$, the behavior of the QDA algorithm implies that \emph{quadratic} decision surfaces are good classifiers. However, because the perceptron and linear SVM algorithms also do (fairly) well, we can conclude that the curvature of the separating surface is small; the surface can be well-approximated by a hyperplane.

The $L=2$ case illustrates why hyperparameter tuning can be necessary. The separability witness (Section \ref{subsec:feasibility_testing}) indicated that $\mathcal{C}_{2}$ was linearly separable, yet \emph{all} of the linear classifiers performed poorly under their default hyperparameters. Hyperparameter tuning improved accuracy, sometimes substantially so.

The LDA algorithm performs poorly, even after hyperparameter tuning. This implies that the assumption of isotropically Gaussian-distributed data is a bad one. Of course, there's no \emph{a priori} reason to expect GST data sets under coherent and stochastic noise have this property, so the poor performance of the LDA algorithm is not too surprising in that regard.

Unsurprisingly, the \emph{linear} classifiers continue to perform poorly on $\mathcal{C}_{1}$, even with hyperparameter tuning. Given that $\mathcal{C}_{1}$ \emph{isn't} linearly separable, this result is to be expected. However, through the use of \emph{feature engineering} -- creating new features out of existing ones -- we can make $L=1$ GST data learnable using linear classifiers.

In the following subsections, we restrict our attention to $\mathcal{C}_{1}$. In the next subsection, we use dimensionality reduction techniques to probe the structure of $\mathcal{C}_{1}$. The insights about this structure inform the design of new feature maps in Section \ref{subsec:feature_engineering} that take $L=1$ GST data sets and map them into a feature space where coherent and stochastic noise becomes linearly separable (Section \ref{subsec:feature_engineering_separable}).

\subsection{Probing the structure of $\mathcal{C}_{1}$ by dimensionality reduction}
\label{sec:feature_space_geom}
\label{subsec:dim_reduction}

The feature vectors in $\mathcal{C}_{1}$ lie in a 92-dimensional feature space. Reasoning about the geometric properties of the data is hard, because those properties cannot be easily visualized. To visualize the structure of high-dimensional data, \emph{dimensionality reduction} techniques are used to embed data from a $(d>>1)$-dimensional feature space into a $(k<<d)$-dimensional space. In this section, we use two dimensionality reduction techniques to examine $\mathcal{C}_{1}$: principal component analysis and multidimensional scaling. We eschew a commonly-used technique, \emph{t-stochastic neighborhood embedding} \cite{VanDerMaaten2008}, because it changes the geometric structure of the embedded data.

\emph{Principal component analysis} (PCA) \cite{Jolliffe2002,Hotelling1933,Pearson1901} is a technique that projects the data onto the directions along which it maximally varies. These directions (the \emph{principal components}) are the eigenvectors of the estimated covariance matrix $\hat{\Sigma}$:
\begin{equation}
\hat{\Sigma} = \sum_{j=1}^{E}\sigma^{2}_{j} \mathbf{e}_{j} \mathbf{e}_{j}^{T},~\text{where}~\hat{\Sigma}\mathbf{e}_{j} = \sigma^{2}_{j} \mathbf{e}_{j}.
\end{equation}
The number of eigenvectors $E$ is less than or equal to $\min(d,N)$, and $\sigma_{j}^{2}$ is the variance of the data along $\mathbf{e}_{j}$. The principal components can be used to define a projector $\Pi_{k}$ from $\mathbb{R}^{d}$ to $\mathbb{R}^{k}$:
\begin{align}
\label{eq:pca_eq}
\mathbf{f}_{j}&\rightarrow\mathbf{y}_{j} \equiv \Pi_{k}[\mathbf{f}_{j}] =\sum_{j=1}^{k}\mathbf{e}_{j}(\mathbf{e}_{j}\cdot\mathbf{f}_{j}).
\end{align}
The dimension of the projector, $k$, is a hyperparameter of this algorithm. In practice, $k < E$, because only the principal components that have \emph{large} eigenvalues are kept in Equation \eqref{eq:pca_eq}: if $\sigma_{j} \sim 0$, the principal component $\mathbf{e}_{j}$ is an ``uninformative" direction.

\emph{Multdimensional scaling} (MDS) \cite{I2005,Torgerson1958,Kruskal1964,Shepard1962a,Shepard1962}  is another approach to dimensionality reduction. It defines  an ideal embedding as one that preserves, as much as possible, all pairwise distances between the feature vectors. So the MDS embedding is a solution to a particular optimization problem. Let $d_{mn}$ denote the distance (or more generally, the dissimilarity) between the feature vectors $\mathbf{f}_{m}$ and $\mathbf{f}_{n}$. Given the set of all pairwise distances $\{d_{mn}\}_{m,n=1}^{N}$, the MDS embedding is
\begin{equation}
\label{eq:mds_opt}
\{\mathbf{y}_{j}\}_{j=1}^{N} = \underset{ \mathbb{R}^{k}}{\text{argmin}}\sum_{m,n=1}^{N}\left(||\mathbf{x}_{m} - \mathbf{x}_{n}|| - d_{mn}\right)^{2},
\end{equation}
where $\mathbf{y}_{j}$ is the embedded version of the feature vector $\mathbf{f}_{j}$. We take the dissimilarity measure to be the Euclidean distance: $d_{mn} = ||\mathbf{f}_{m} - \mathbf{f}_{n}||_{2}$.

\begin{figure}
\includegraphics[width=\columnwidth]{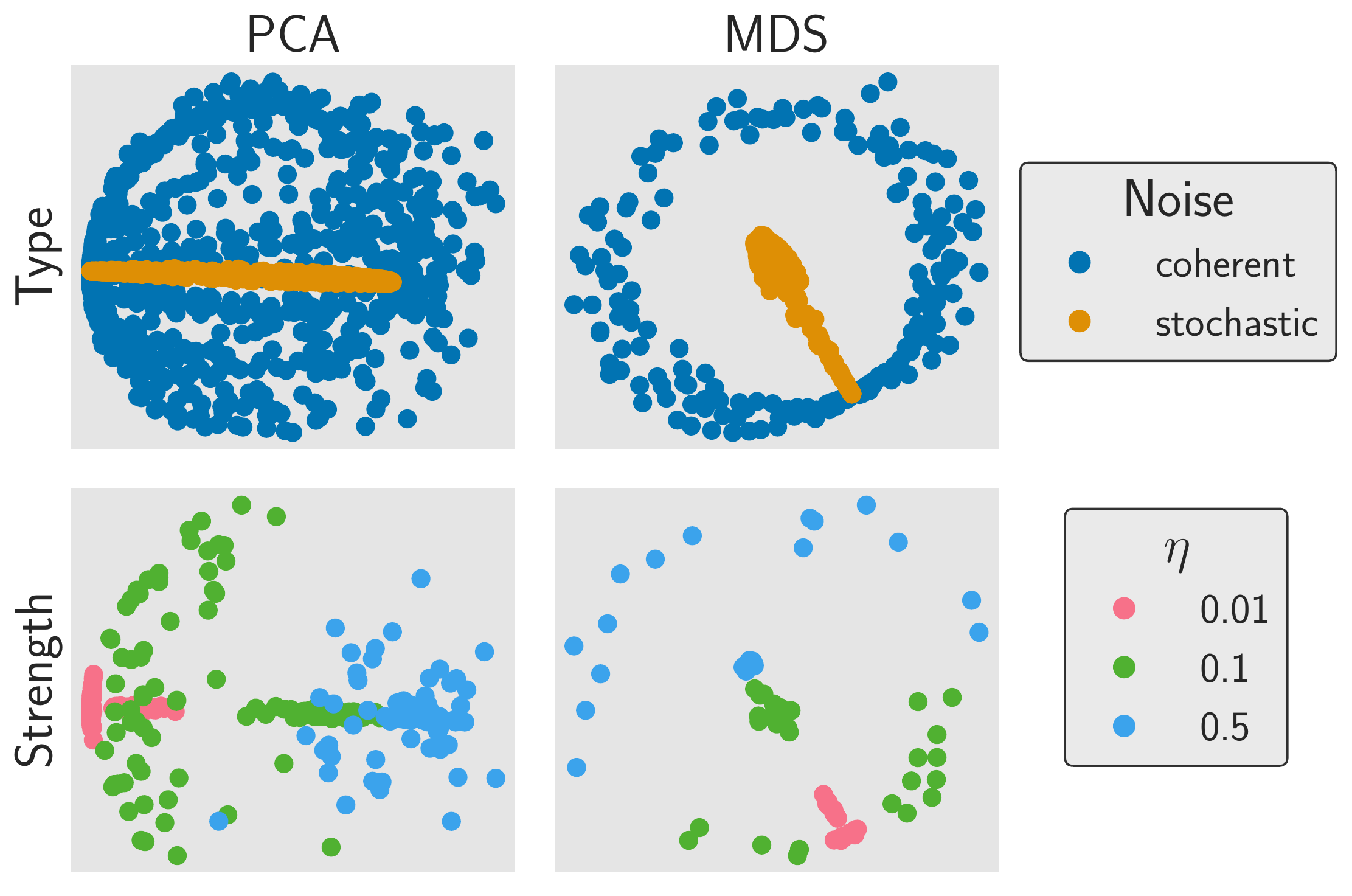}
 \caption{\textbf{2-dimensional embeddings of $\mathcal{C}_{1}$}. \textbf{Both}: $\mathcal{C}_{1}$ appears to have a structure similar to a high-dimensional radio dish. \textbf{Top}: The embedded feature vectors for stochastic noise appear to be ``surrounded" by those for coherent noise. \textbf{Bottom}: As $\eta \rightarrow 0$, the feature vectors for coherent and stochastic noise approach one another, which is to be expected, as at $\eta = 0$, the ``noisy" gate set generating the feature vectors is the ideal (noiseless) one.}
\label{fig:pca_mds}
\end{figure}

Figure \ref{fig:pca_mds} plots the $k=2$-dimensional embeddings of $\mathcal{C}_{1}$ using PCA or MDS. The top row colors the embedded points by noise \emph{type}, and the bottom row colors them by noise \emph{strength}. Both plots indicate that $\mathcal{C}_{1}$ bears some resemblance to a high-dimensional radio dish: the ``bowl" of the dish is formed by feature vectors arising from coherent noise, while the ``antenna'' comes from feature vectors arising from stochastic noise. This picture illustrates \emph{why} $\mathcal{C}_{1}$ is linearly inseparable: the two parts of a radio dish cannot be separated using a hyperplane.

These embeddings are low-dimensional approximations to high-dimensional feature spaces, leading to the question ``Are these `radio dishes' real?" A simple argument suggests the answer is ``mostly yes." The circuits used for $L=1$ GST generate a feature vector that depends on the gate set in an almost linear fashion \footnote{Note that in process tomography, the feature vector would be \emph{exactly} linear in the process.}. The deviations from linearity are small, so the geometry of $\mathcal{C}_{1}$ is similar to the structure of the underlying gate sets generating the feature vectors.

This structure can be understood by mapping the gate set to a quantum state using the Choi-Jamio\l{}kowski isometry \cite{Jamiokowski1972}. A gate set is the direct sum of the constituent gates (a linear operation), and the Choi-Jamio\l{}kowski isometry is a linear map from gates (channels) to quantum states. Hence, a linear map exists taking gate sets into quantum state space. A gate affected by purely coherent noise maps to a pure state, and a gate affected by purely stochastic noise maps to a mixed state.  Just as pure states form the extremal points of quantum space, ``enveloping" the mixed states, gates affected by purely coherent noise  ``envelop" those affected by purely stochastic noise. This behavior is analogous to that observed with $\mathcal{C}_{1}$, although the exact structure may not be comparable. (Again, there are small deviations from a linear relationship between feature vectors and gate sets for $L=1$ GST.) This strongly suggests that the structure present in $\mathcal{C}_{1}$ is genuine, and not an artifact of our simulations.

Given this knowledge about the geometry of $\mathcal{C}_{1}$, we can use \emph{feature engineering} to change the geometry of the feature vectors. In particular, $\mathcal{C}_{1}$ can be ``unrolled" in such a way that \emph{linear} classifiers achieve high accuracy. We introduce two new feature maps in the next subsection, and Section \ref{subsec:feature_engineering_separable} shows that the perceptron and linear SVM algorithms learn a high-accuracy separating hyperplane in these new feature spaces.

\subsection{Overcoming linear inseparability of $\mathcal{C}_{1}$ using feature engineering}
\label{subsec:feature_engineering}

Linear classifiers on the base feature space performed poorly on $\mathcal{C}_{1}$, because the natural separating surface has curvature. A linear classifier can learn a curved decision surface in the base feature space, but only if they are given access to new features. That is, by adding new features to the base feature space, and ``unrolling" the radio dish structure present in $\mathcal{C}_{1}$, then linear classifiers could potentially achieve accuracies comparable to intrinsically nonlinear algorithms.

As noted in Section \ref{sec:ml_qcvv_framework}, adding new features to a base feature space is called \emph{feature engineering}. We do feature engineering here, and define two new feature maps. Both add new components to the feature vectors on top of the base features defined by $\phi$, thereby enlarging the feature space.

Let $\mathcal{D}$ be a GST data set with outcome frequencies $(f_{1}, f_{2}, \cdots, f_{d})$. Define two new feature maps $\phi_{\mathrm{SQ}}$ and $\phi_{\mathrm{PP}}$ as 
\begin{align}
\begin{split}
\phi_{\mathrm{SQ}}&: \mathcal{D} \rightarrow \mathbf{f} \in [0,1]^{2d}\\
\phi_{\mathrm{SQ}}(\mathcal{D}) &=  \left( \underset{j}{\oplus} f_{j}\right) \oplus \left( \underset{j}{\oplus} f_{j}^{2}\right),
\end{split}
\end{align}
and
\begin{align}
\begin{split}
\phi_{\mathrm{PP}}&: \mathcal{D} \rightarrow \mathbf{f} \in  [0,1]^{d(d+3)/2}\\
\phi_{\mathrm{PP}}(\mathcal{D}) &=  \left( \underset{j}{\oplus} f_{j}\right)\oplus \left(\underset{j<k}{\oplus} f_{j}f_{k}\right).
\end{split}
\end{align}
Note that $\phi_{\mathrm{PP}}$ only adds \emph{unique} pairwise products, since $f_{j}f_{k}$ and $f_{k}f_{j}$ are redundant. As an example of how these maps act, consider a data set $\mathcal{D}$ with outcome probabilities $[f_{1}, f_{2}]$. Then $\phi_{\mathrm{SQ}}(\mathcal{D}) = (f_{1}, f_{2}, f_{1}^{2}, f_{2}^{2}) \in \mathbb{R}^{4}$, and $\phi_{\mathrm{PP}}(\mathcal{D}) = (f_{1}, f_{2}, f_{1}f_{2}, f_{1}^{2}, f_{2}^{2}) \in \mathbb{R}^{5}$.

Both feature maps create new features that are simple polynomial functions of the features in the base feature space, but they affect the feature space in different ways. $\phi_{\mathrm{SQ}}$ adds a quadratic nonlinearity while preserving the coordinate axes, whereas $\phi_{\mathrm{PP}}$ allows quadratic nonlinearity together with rotation of the coordinate axes. If the coordinate axes have significance,  $\phi_{\mathrm{SQ}}$ is preferable, while  $\phi_{\mathrm{PP}}$ is preferable if correlations between different variables are important. Also, note that $\phi_{\mathrm{PP}}$ gives classifiers quadratically more parameters than $\phi_{\mathrm{SQ}}$ and hence, the classifiers may overfit the data more.

\subsection{Feature engineering of $\mathcal{C}_{1}$ enables linear separability}
\label{subsec:feature_engineering_separable}

\begin{figure}
\includegraphics[width=\columnwidth]{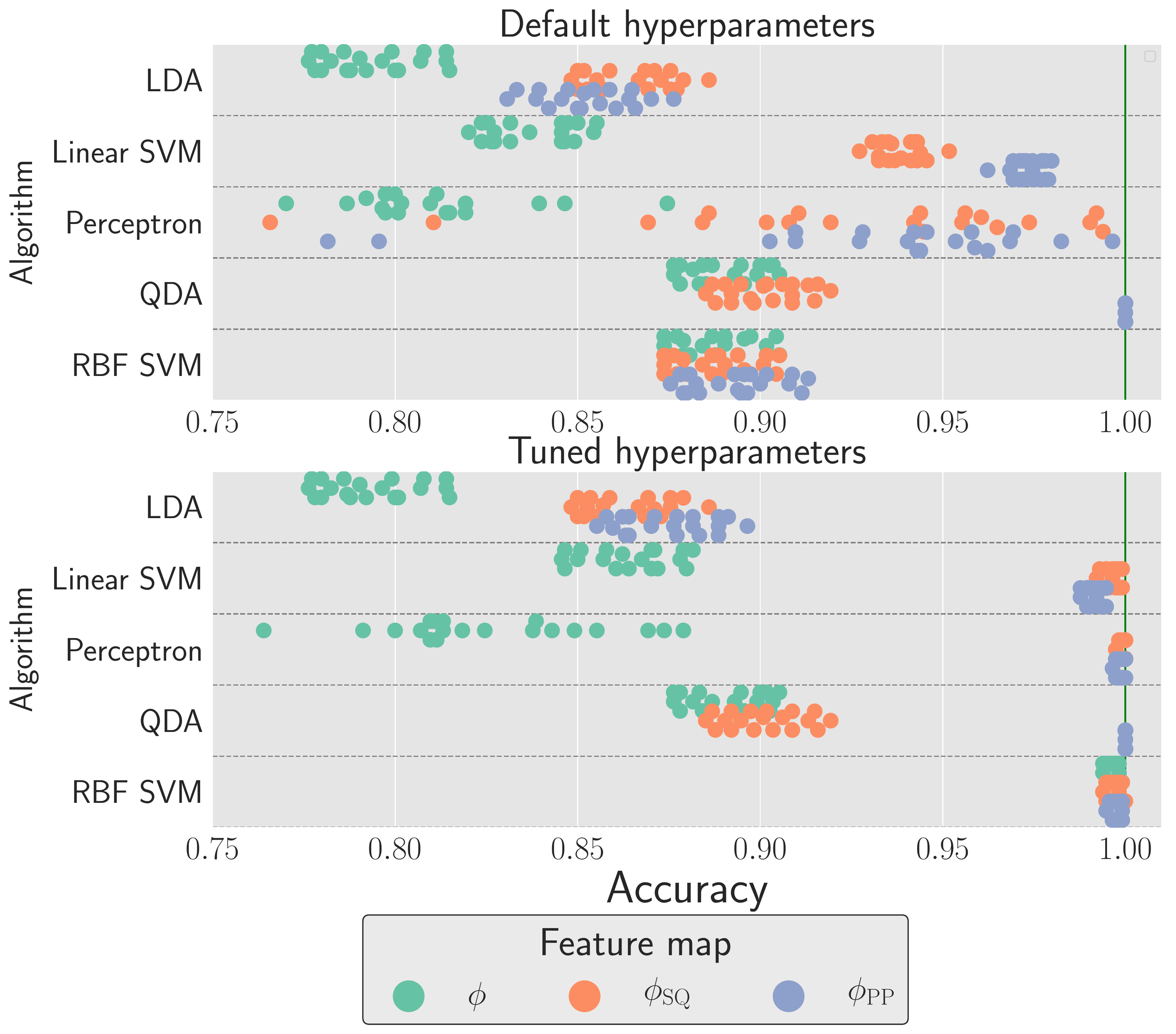}
 \caption{\textbf{Feature engineering boosts accuracy of some classification algorithms ($L=1$)}. Cross-validated accuracies of the algorithms under their default hyperparameters (top) and with tuned hyperparameters (bottom). The performance of the RBF SVM algorithm under the base feature map $\phi$ indicates \emph{nonlinear} separating surfaces are best. The QDA algorithm's performance under the $\phi_{\mathrm{PP}}$ feature map suggests a ``quartic surface" -- a 4$^{\mathrm{th}}$ order polynomial in the circuit outcome probabilities -- is a sufficient amount of nonlinearity. The performance of the linear SVM and perceptron algorithms implies quadratic separating surfaces work well as approximations to the nonlinear surfaces learned by the RBF SVM and/or QDA algorithms.}
\label{fig:cv_2}
\end{figure}

Having defined two new feature maps that should, in principle, enable $L=1$ GST data sets to be classified by linear classifiers, we first check whether the resulting feature vectors are in fact linearly separable. The separability witness (Appendix \ref{app:lp_problem}) indicates that under the action of $\phi_{\mathrm{SQ}}$ and $\phi_{\mathrm{PP}}$, the previously-inseparable $L=1$ GST data sets become linearly separable in the new feature spaces. Because of this, linear classifiers should perform better than they did on $\mathcal{C}_{1}$.  Figure \ref{fig:cv_2} shows feature engineering does boost the performance of the linear classifiers, and that this boost can be increased using hyperparameter tuning. See Figure \ref{fig:cv_engineering} in Appendix \ref{app:hyperparamsweep} for results of the hyperparameter sweep.

In Figure \ref{fig:cv_2}, the performance of the RBF SVM algorithm on the base feature space indicates that \emph{nonlinear} separating surfaces are generally best. However, the amount of nonlinearity required is modest: the performance of the QDA algorithm on $\phi_{\mathrm{PP}}$ feature vectors suggests a ``quartic surface" -- a 4$^{\mathrm{th}}$ order polynomial in the features  -- is a sufficient amount of nonlinearity. Recall that QDA learns a quadratic classification rule, and $\phi_{\mathrm{PP}}$ uses all pairwise products. Therefore, the classification rule is a \emph{quartic} function of the features in the base feature space.

The fact that linear SVM and perceptron algorithms also work well implies quadratic separating surfaces are good approximations to the nonlinear surfaces learned by the RBF SVM and/or QDA algorithms. Therefore, while the curvature of a natural separating surface may deviate from a flat curvature, those deviations can be captured by quadratic surfaces.

As noted in Section \ref{subsec:feasibility_testing}, there is always a risk of under-sampling the number of noise realizations. We again checked for undersampling effects, and did not find any (Table \ref{tab:bestHyperparams} in Appendix \ref{app:hyperparamsweep}).

In summary: \textbf{\emph{feature engineering makes $\mathcal{C}_{1}$ linearly separable, so the accuracy of (some) linear classifiers becomes comparable to that of intrinsically nonlinear algorithms.}}

\subsection{Robustness to finite-sample effects}
\label{sec:finite_sampling}

The tests and analysis done up to this point have been performed in the \emph{exact-sampling} limit. However, real QCVV data sets have finite-sample fluctuations.  Finite sampling means the outcome frequency of a circuits run on a QIP will \emph{fluctuate} around the exact outcome probabilities by $\mathcal{O}(1/\sqrt{N_{\mathrm{samples}}})$. If the amount of fluctuation is sufficiently small, then classifiers trained on fluctuation-free data should still be able to reliably classify a feature vector with finite-sample effects. The reason is simple: suppose a hyperplane $H$ had been learned by training a classification algorithm on fluctuation-free data.  $H$ would also do well in classifying finite-sample data, provided the statistical fluctuations are less than the geometric margin $M_{H}$.

$M_{H}$ is a property of both the hyperplane $H$ and the data it separates [recall Equation \eqref{eq:geom_margin}]. Hyperplanes with larger margins are more robust to finite-sample fluctuations, so a hyperplane that \emph{maximizes} the margin is preferable. Consequently, the SVM is the ideal ML algorithm for learning a separating hyperplane that would later be used to classify finite-sample data.

To demonstrate the robustness of the linear SVM algorithm, we restrict our attention to $\mathcal{C}_{1}$. We do so for two reasons: the $L=1$ GST experiment design is the simplest one possible, and because we can examine the interplay of feature engineering, finite-sample fluctuations, and classification accuracy.

To evaluate how the hyperplane learned by the SVM algorithm performs in the presence of finite-sample effects, we first start by training a soft-margin SVM ($C=10^{5}$) on the fluctuation-free data in the feature spaces defined by $\phi_{\mathrm{SQ}}$ and $\phi_{\mathrm{PP}}$. This training yields a separating hyperplane $H=(\bs{\beta}, \beta_{0})$. This fixed hyperplane is then used to classify noisy data sets. We generate these data sets by taking the fluctuation-free feature vectors and adding finite-sampling noise by simulating independent $\mathrm{Binomial}(N_{\mathrm{samples}}, \mathbf{f}_{j})$ random variables for each component of the feature vector.

After the noisy version of $\mathbf{f}$ is generated, then the either of the maps $\phi_{\mathrm{SQ}}$ or $\phi_{\mathrm{PP}}$ is applied. The resulting data collection is then mean-standardized (see Section \ref{subsec:data_set_descrip}) using the estimated mean of the fluctuation-free data. (Mean-standardization doesn't distort the margin of $H$, and also helps with the training of the SVM.) Finally, $\mathbf{f}'$ is classified using $H$, and the label assigned by the decision surface is compared to the actual label. We use 50 independent realizations of the finite-sample noise for each value of $N_{\mathrm{samples}}$.

\begin{figure}
\includegraphics[width=\columnwidth]{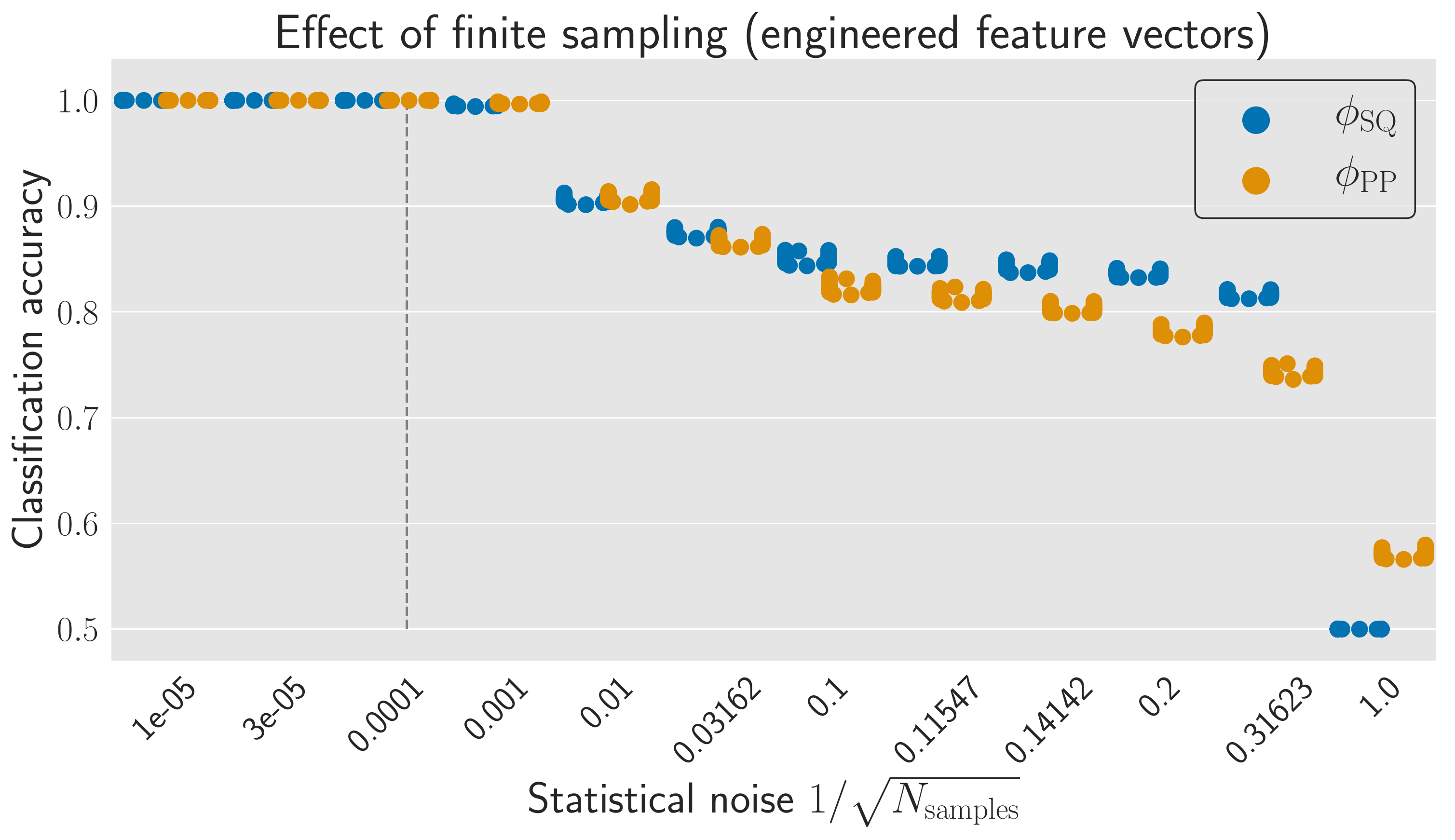}
 \caption{\textbf{Classification accuracy under finite-sampling effects using feature engineering ($L=1$)}.
A soft-margin linear SVM ($C=10^{5}$) was trained using \emph{noiseless} feature-engineered feature vectors, and then the accuracy of the hyperplane it learned was evaluated using noisy versions of that data. The geometric margin of the hyperplane learned by the SVM algorithm was comparable between the two feature engineering maps, and is indicated by the dashed vertical line. Once the statistical noise is less than the geometric margin, the accuracy goes to 1.}
\label{fig:finite_sampling_engineered}
\end{figure}

Results of this test are shown in Figure \ref{fig:finite_sampling_engineered}. The vertical grey line shows the margin of the hyperplane learned by the SVM algorithm \footnote{$M_{H}$ is more-or-less the same regardless of whether the SVM was trained in the feature space defined by $\phi_{\mathrm{SQ}}$ or $\phi_{\mathrm{PP}}$}.  Once the statistical noise is less than the margin of the hyperplane, classification accuracy increases to 1.
\textbf{\emph{This test confirms the intuition that maximum-margin hyperplanes are more robust for classifying noise in the presence of finite-sample effects.}}

The two feature engineering maps considered here add terms of the form $\hat{p}_{j}^{2}$ or $\hat{p}_{j}\hat{p}_{k}$ to the original feature vector. Simple algebra shows that the fluctuations in $\hat{p}_{j}^{2}$ and $\hat{p}_{j}\hat{p}_{k}$ both go as $\mathcal{O}(1/\sqrt{N_{\mathrm{samples}}})$. Thus, adding these components doesn't make the engineered feature vector \emph{more} sensitive to finite-sample effects. However, it doesn't make the feature vector \emph{less} sensitive, either. An open question is whether there are other feature engineering approaches that would still enable linear classifiers to perform well on $\mathcal{C}_{1}$, while decreasing the sensitivity of the feature vector to finite-sample noise.

\section{Discussion and conclusions}
\label{sec:conclusions}

As quantum computing enters the NISQ era, determining what NISQ processors are useful for becomes increasingly important, and to extend their computational capability. More computational power requires longer circuits, which demand lower error rates \footnote{For some metrics of computational utility -- such as the \emph{quantum volume} \cite{Cross} -- lowering the error rate past some effective threshold doesn't improve the metric.}. If the QIP's error rate is $\epsilon$, a circuit whose size exceeds $\mathcal{O}(1/\epsilon)$ will most likely output the wrong answer. Characterizing a processor helps to improve its performance, but characterizing NISQ processors comes with its own set of unique challenges. New QCVV techniques are needed.

What we have demonstrated here is that machine learning algorithms can help develop new QCVV techniques, as discussed in Section \ref{subsec:ml_helps}. ML algorithms don't model the underlying complexity of a QIP in detail. Instead, they operate from the premise that a QCVV technique simply needs to approximate the functional relationship between data and a QIP property. ML algorithms excel at learning approximations to functions, so they can help automate that task.

This doesn't mean that QCVV practitioners are unnecessary. Expertise is still needed to propose relevant QIP properties to be characterized, or devise an appropriate experiment design. Experts are needed to choose between algorithms, evaluate their performance, write code, interpret results, etc. ML algorithms \emph{augment} the expertise of QCVV practitioners, rather than \emph{replacing} it.

Domain-specific expertise helps can guide wise choices for the feature map, suggest good feature engineering techniques, or help corroborate intuition gleaned from purely ML-based approaches. As we saw in Section \ref{subsec:feature_engineering}, knowing how coherent and stochastic noise affects gate sets helped confirm the intuition that $\mathcal{C}_{1}$ looked like a radio dish, and provided a measure of confidence the dimensionality reduction techniques we used were actually showing us something about the structure of $\mathcal{C}_{1}$.

At the same time, QCVV practitioners using ML will need to be conversant in the language, methodologies, and vagaries of ML algorithms. Because there are a plethora of ML algorithms for a wide variety of tasks, QCVV practitioners will need to make informed and prudent judgements about which algorithms to deploy. Here, supervised binary classifiers were the right choice because the characterization task involved inferring a particular binary property of the QIP, and because generating synthetic training data was easy.

Our work showed that some of these algorithms can successfully learn a QCVV technique for distinguishing single-qubit coherent and stochastic noise. Their success depended strongly on the algorithm's own hyperparameters and the feature map used to embed experimental data into a feature space.

The success in deploying classification algorithms suggests that \emph{regression} algorithms could be used as well to estimate the coherence of the noise. One of the advantages in using ML is that, with a sufficiently rich and well-described data set, different algorithms can be deployed using the \emph{same} data, to solve \emph{different} tasks. For example, if the training data contained feature vectors, noise labels, and a measure of the noise strength, then both classification and regression algorithms could be trained.

We generated synthetic experimental data using computer simulations of noisy single-qubit QIPs. Scaling these simulations up will be difficult. Simulation will become harder once quantum supremacy has been shown, or quantum advantage attained, simply because of the number of qubits to be simulated. Even if the noise acts only on a sparse number of qubits, ensuring the data collection is representative could be challenging. One possible way to remedy these difficulties is to use well-calibrated devices to generate training data by, e.g., deliberately injecting specific kinds of noise into the QIP when it is running circuits.

This work used machine learning algorithms as part of a data processing pipeline. The other major task of a QCVV practitioner is coming up with a good \emph{experiment design} whose outcome probabilities are useful for inferring the property of interest. We alluded to the question of machine-learned experiment design Section \ref{subsec:ml_helps}. Experiment design is a non-trivial problem, which is why this work borrows the experiment design for gate set tomography. We see at least two ways that ``machine-learned experiment design" could be pursued. The first is to use ML algorithms for \emph{feature selection} by selecting, out of a candidate set of circuits, a smaller subset that is useful for characterizing a property. Another approach is to use  \emph{reinforcement learning} (RL) \cite{Sutton} to construct QCVV experiment designs from scratch, as RL has already been used for other quantum computing tasks \cite{Fosel2018}, and classical experiment design \cite{Gatti2015,Gatti2013}
~\\~\\
As QIPs advance and the rate of their advancement increases, QCVV theorists are faced with the challenge of developing increasingly-powerful characterization techniques in ever-shorter timeframes. We hope and anticipate that leveraging ML algorithms can help them do so.

\section{Acknowledgements}

The authors are grateful for those who write and maintain code for the following software packages: cvxpy \cite{Diamond2016, Agrawal2018}, Jupyter \cite{Perez}, matplotlib \cite{Hunter2007},  NumPy \cite{VanDerWalt2011}, pandas \cite{mckinney2010}, pyGSTi \cite{Nielsen2017}, Python 2.7 \cite{vanRossum}, scikit-learn \cite{Pedregosa2011}, SciPy \cite{Oliphant2007a}, and seaborn \cite{Waskom2016}. TLS thanks Giacomo Torlai and Justin C. Johnson for feedback on earlier drafts of this paper. TLS acknowledges the Center for Quantum Information and Control (CQuIC) at the University of New Mexico, where the bulk of this work was carried out during his PhD. The authors also thank various internal reviewers for their thoughtful and detailed feedback.

Sandia National Laboratories is a multimission laboratory managed and operated by National Technology \& Engineering Solutions of Sandia, LLC, a wholly owned subsidiary of Honeywell International Inc., for the U.S. Department of Energy's National Nuclear Security Administration under contract DE-NA0003525. This paper describes objective technical results and analysis. Any subjective views or opinions that might be expressed in the paper do not necessarily represent the views of the U.S. Department of Energy or the United States Government. Contributions to this work by NIST, an agency of the US government, are not subject to US copyright. Any mention of commercial products is for informational purposes only, and does not indicate endorsement by NIST.
\newpage 

\appendix

\section{Numerically simulating purely coherent or purely stochastic noise}
\label{app:noisesim}
Both coherent and stochastic noise (as defined in Section \ref{sec:qipnoise}) are \emph{Markovian}. The most general Markovian, continuous-time dynamics of a $d$-dimensional quantum system is described by the Lindblad master equation \cite{Breuer2007,Lindblad1976,Gorini1976}:

\begin{align}
\begin{split}
\label{eq:lindblad_eq}
\dot{\rho}&= -\frac{i}{\hbar}[H(t), \rho] + \sum_{j,k=1}^{d^2 -1}h_{jk}(t)\left[A_{j}\rho A_{k}^{\dagger}- \frac{1}{2}\{A^{\dagger}_{k}A_{j},\rho\}\right]\\
&\equiv \mathcal{L}[\rho],
 \end{split}
\end{align}
where $H(t)$ is the Hamiltonian and the set of operators $\{A_{j}\}$ forms an orthonormal basis for Hermitian matrices. The first term in the Lindblad equation generates unitary dynamics, while the second generates noisy dynamics such as dephasing, amplitude damping, or bit-flip noise.

In our simulations, we make a simplification to Equation \eqref{eq:lindblad_eq} by assuming $H$ and the set $\{A_{j}\}$ are all \emph{time-independent}.  Under this simplification, Equation \eqref{eq:lindblad_eq} is \emph{time-invariant}.
For the time-invariant Lindblad equation, the evolution of $\rho$ can be written as $\rho(t) = e^{\mathcal{L}t}[\rho(0)]$. In the circuit model, the updates are discrete (from one timestep of the circuit to the next); the corresponding quantum channel is $\mathcal{E}[\rho] = e^{\mathcal{L}}[\rho]$.

As noted in Section \ref{sec:qipnoise}, we imagine that \emph{regardless} of when the  Hamiltonian generating an ideal gate is turned on when the QIP performs the gate, \emph{the exact same set of noise operators are turned on at the same time}.

For purely coherent noise, the dynamics generated by the Lindblad equation must be purely unitary, so the jump coefficients $h_{jk}$ are all 0:
\begin{equation}
\dot{\rho} = -\frac{i}{\hbar}[H, \rho].
\end{equation}

Letting $H_{0}$ be the generator of the ideal gate, the total Hamiltonian can be written as $H = H_{0} + H_{e}$, with $H_{e}$ as the error. Each parameter in $H_{0}$ is usually associated with some external classical control field:
\begin{equation}
H_{0} = \sum_{j=X,Y,Z}c_{j}\sigma_{j},
\end{equation}
where the coefficient $c_{j}$ controls the evolution of the qubit about the $\sigma_{j}$ axis. Thus, a natural choice for $H_{e}$ is to draw it from the Gaussian unitary ensemble:
\begin{equation}
H_{e} = a\sigma_{X}+b\sigma_{Y}+c\sigma_{Z}~~~~a,b,c\sim \mathcal{N}(0,\eta^{2}).
\end{equation}
For this model of purely coherent noise, each control $c_{j}$ is subject to independent and identically distributed noise with mean zero, variance $\eta^{2}$, and \emph{is constant in time}.

For purely stochastic noise on a single qubit, there should be no unitary dynamics except the action of $H_{0}$:
\begin{equation}
\label{eq:purely_stochastic}
 \dot{\rho} = -\frac{i}{\hbar}[H_{0}, \rho] + \sum_{j,k=1}^{3}h_{jk}\left[A_{j}\rho A_{k}^{\dagger}+\frac{1}{2}\{A_{k}^{\dagger}A_{j},\rho\}\right].
\end{equation}
Defining a superoperator $\mathcal{S}$ as
\begin{equation}
\mathcal{S}[\rho] = \sum_{j,k=1}^{3}h_{jk}\left[A_{j}\rho A_{k}^{\dagger}+\frac{1}{2}\{A_{k}^{\dagger}A_{j},\rho\}\right],
\end{equation}
the noisy channel can be written as
\begin{equation}
\mathcal{E} = e^{\mathcal{H}_{0} + \mathcal{S}}[\rho].
\end{equation}

We take the jump operators to be the Pauli matrices: $A_{j} = \sigma_{j}$. To simulate stochastic noise, we generate the coefficient matrix $h$  as
\begin{align}
\begin{split}
h_{jk} &= (S^{-1}DS)_{jk}\\
D &= \text{diag}(a,b,c)\\
a,b,c &\sim |\mathcal{N}(0, \eta^{2})|,
\end{split}
\end{align}
where the columns of $S$ form a randomly-chosen basis for $\mathbb{R}^{3}$. The variables $a,b,c$ are drawn from a \emph{folded normal} distribution.

We call a particular randomly-generated value for $H_{e}$ or the coefficient matrix $h$ a \emph{realization} of the corresponding noise type. Again, we emphasize that because we are using a time-invariant master equation, once a realization has been generated, then any time $G_{0}$ occurs in the compilation of a given unitary, it is replaced by the \emph{same} noisy version. Commonly when simulating these noise types, different realizations of the noise are generated \emph{each time}  $G_{0}$ occurs. We do not use this approach.

For both coherent and stochastic noise, as $\eta \rightarrow 0$ both $H_{e}$ and $h$ vanish. (That is, at $\eta =0$, the channel is noiseless.) The role of $\eta$ can be formalized as follows. For purely coherent noise, the average over all realizations of the error Hamiltonian $H_{e}$ is $0$:
\begin{equation}
\langle H_{e} \rangle = \langle a \rangle \sigma_{X} +\langle b \rangle \sigma_{Y} + \langle c \rangle \sigma_{Z} = 0.
\end{equation}
The \emph{variance} of the error Hamiltonian is non-zero, however:
\begin{align}
\begin{split}
\left(\Delta H_{e}\right)^{2} &= \langle H_{e}^{2}\rangle - \left(\langle H_{e}\rangle\right)^{2}\\
&=\left \langle \left(a \sigma_{X} + b  \sigma_{Y} +  c \sigma_{Z} \right)^{2}\right\rangle\\
& = 3\eta^{2}\mathcal{I},
\end{split}
\end{align}
because the noise realizations $a,b,c$ are i.i.d. random variables. Therefore, the fluctuations in the noise are isotropic, and their magnitude is proportional to $\eta^{2}$.

For purely stochastic noise, consider the expected value of the coefficients $h_{jk}$ in Equation \eqref{eq:purely_stochastic}. To compute $\langle h_{jk}\rangle$, it's important to note the randomly-chosen basis for $\mathbb{R}^{3}$ is \emph{independent} of the random variables $a,b$, and $c$, which means
\begin{align}
\begin{split}
\langle h_{jk}\rangle &= \langle (S^{-1}DS)_{jk}\rangle = \sum_{qr}\langle (S^{-1})_{jq}D_{qr}S_{rk}\rangle\\
&= \sum_{r} \langle (S^{-1})_{jr}S_{rk}\rangle  \underbrace{\langle D_{rr} \rangle}_{\eta \sqrt{2\pi}} \propto \eta.
\end{split}
\end{align}
Unlike the case of coherent noise, the expected value of the noise generator is non-zero.
The expected value of $h_{jk}^{2}$ is
\begin{align}
\begin{split}
\langle h_{jk}^{2}\rangle &=\left \langle  \sum_{qr} (S^{-1})_{jq}D_{qr}S_{rk} \sum_{q'r'}(S^{-1})_{jq'}D_{q'r'}S_{r'k}\right\rangle\\
&=\left \langle  \sum_{qq'rr'} (S^{-1})_{jq}S_{rk} (S^{-1})_{jq'}S_{r'k}D_{qr}D_{q'r'}\right\rangle\\
&=\sum_{qq'} \langle(S^{-1})_{jq}S_{qk} (S^{-1})_{jq'}S_{q'k}\rangle \langle D_{qq}D_{q'q'}\rangle\\
&=\sum_{q\neq q'} \langle(S^{-1})_{jq}S_{qk} (S^{-1})_{jq'}S_{q'k}\rangle\langle D_{qq}\rangle\langle D_{q'q'}\rangle\\
&+\sum_{q=q'} \langle(S^{-1})_{jq}S_{qk}]^{2}\rangle \langle D^{2}_{qq}\rangle \propto \eta^{2}.
\end{split}
\end{align}
Similar to the case of purely coherent noise, as $\eta\rightarrow 0$, the noise terms more and more tightly concentrate around 0.

\section{Testing for linear separability as a linear programming problem}
\label{app:lp_problem}
There is a straightforward test for determining whether a given data set is linearly separable using \emph{linear programming}. Various formulations of this problem have been put forth \cite{Roychowdhury1995, Basu2006}. Our formulation is simple -- if the data is not linearly separable, the corresponding linear programming problem \emph{should not} be feasible. The derivation below closely follows the line of reasoning presented in \cite{Vogler2014}.

Consider two data sets $A = \{\mathbf{a}_{1}, \cdots \mathbf{a}_{n}\}$ and  $B = \{\mathbf{b}_{1}, \cdots \mathbf{b}_{m}\}$ where $\mathbf{a}_{j}, \mathbf{b}_{k} \in \mathbb{R}^{d}$. If $A$ and $B$ are linearly separable, there exists a hyperplane $H$ defined by a normal vector $\bs{\beta}$ and scalar offset $\beta_{0}$ such that
\begin{equation}
\bs{\beta}\cdot\mathbf{a}_{j} > \beta_{0},~\bs{\beta}\cdot\mathbf{b}_{k} < \beta_{0}~~\forall~j,k.
\end{equation}
Note this implies that the label for $A$ is $1$, while the label for $B$ is $-1$, as the classification rule implied by this hyperplane is $\mathbf{f} \rightarrow \text{sign}\left[\bs{\beta}\cdot \mathbf{f} - \beta_{0}\right]$.

Multiplying the first inequality by $-1$ flips the sign of the inequality, resulting in the following conditions:
\begin{align}
\begin{split}
-\bs{\beta}\cdot\mathbf{a}_{j} + \beta_{0} &< 0,\\
\bs{\beta}\cdot\mathbf{b}_{k} - \beta_{0} &< 0~~\forall~j,k.
\end{split}
\end{align}
These two inequalities can be formulated as a single matrix inequality by defining a new vector $\tilde{\bs{\beta}} = (\bs{\beta}, \beta_{0})$ and new feature vectors $\mathbf{a}_{j} \rightarrow (-\mathbf{a}_{j}, 1)$ and $\mathbf{b}_{k} \rightarrow (\mathbf{b}_{k}, -1)$. In terms of these variables, the inequality constraints become
\begin{equation}
\begin{pmatrix}-\mathbf{a}_{1}^{T}, 1 \\ -\mathbf{a}^{T}_{2}, 1 \\ \vdots \\ -\mathbf{a}^{T}_{n}, 1 \\ \mathbf{b}_{1}, -1 \\ \vdots \\ \mathbf{b}_{m}, -1\end{pmatrix}\tilde{\bs{\beta}} < \mathbf{0}.
\end{equation}

Defining $D$ to be the $(n + m) \times (d  + 1)$ matrix above, it follows that \textbf{if $A$ and $B$ are linearly separable, there exists a $\tilde{\bs{\beta}}$ such that $D\tilde{\bs{\beta}} < \mathbf{0}$}.
Determining if any such $\tilde{\bs{\beta}}$ exists can be done by recasting the problem as a \emph{linear programming} problem:
\begin{align}
\begin{split}
\label{eq:lp_problem}
\tilde{\bs{\beta}} &= \underset{\mathbf{x} \in \mathbb{R}^{N+1}}{\text{argmin}}~~\mathbf{0} \cdot \mathbf{x}\\
 & \text{s.t.}~D\mathbf{x} < \mathbf{0}.
\end{split}
\end{align}
If $A$ and $B$ are linearly separable, there exists at least one feasible solution to this problem. If Equation \eqref{eq:lp_problem} is feasible, then a convex optimization problem should be able to solve it. Suppose, though, that a solver \emph{didn't} find a solution -- how could we tell that this happened because the problem was in fact \emph{infeasible}, and not because the solver terminated or crashed prematurely?

Equation \eqref{eq:lp_problem} defines a \emph{strict} system of inequalities. For any such problem, a theorem of \emph{alternatives} \cite[Example 2.21]{Boyd} gives the conditions under which it is feasible. Given a system of strict inequalities
\begin{equation}
A\mathbf{x} < \mathbf{b},
\end{equation}
the system is \emph{infeasible} if, and only if, there exists $\mathbf{y}$ satisfying
\begin{equation}
\mathbf{y}\neq 0,~\mathbf{y} > 0,~A^{T}\mathbf{y} = 0,~\mathbf{y}\cdot \mathbf{b} \leq 0.
\end{equation}
Thus, suppose that in the course of solving Equation \eqref{eq:lp_problem}, a solver doesn't find a solution. We can easily check whether the problem is infeasible by demonstrating a solution to the following problem:
\begin{equation}
\mathbf{y}\neq 0,~\mathbf{y} > 0,~D^{T}\mathbf{y} = 0.
\end{equation}
This system is nothing more than the \emph{dual problem} of the original LP (Equation \eqref{eq:lp_problem}). Thus, determining whether a given data set is linearly separable gives rise to a set of alternatives: \emph{if the primal problem is feasible, the data is linearly separable, whereas if the dual problem is feasible, the data is linearly inseparable; what's more, the primal and dual problems cannot both be feasible simultaneously.}
This allows us to certify that a given data set is inseparable, by forming the dual problem and finding a solution.

\textbf{\emph{In sum, $A$ and $B$ are separable if, and only if, the optimization problem \eqref{eq:lp_problem} is feasible.}}

\section{Hyperparameter sweeps}
\label{app:hyperparamsweep}

\begin{table*}
\begin{center}
\begin{tabular}{|l|l|c|p{2.8cm}|p{3cm}|}
\hline
  Algorithm & $\phi$ &  Hyperparameter value &  Mean accuracy ($K=20$ cross-validation)  & ``Hero test" accuracy  \\\hline
\midrule
        LDA &     PP &        $10^{-1}$ &       0.87 & 0.87\\\hline
        LDA &     SQ &        $10^{-5}$ &       0.86 & 0.867\\\hline
 Linear SVM &     PP &     75 &       0.991 & 0.97 \\\hline
 Linear SVM&     SQ &      250 &       0.997 & 0.994 \\\hline
 Perceptron &     PP &       100 &       0.999 &0.997\\\hline
 Perceptron &     SQ &       100 &       0.9996 & 0.9994
\\\hline
       QDA &     PP &               0 &       1.0 & 1.0\\\hline
       QDA &     SQ &               0 &       0.90 & 0.90\\\hline
       RBF SVM & PP &$C=20, \gamma=.01$ & 0.998& 0.9978\\\hline
       RBF SVM & SQ & $C=10, \gamma=1$ & 0.997&0.97\\\hline
\bottomrule
\end{tabular}
\caption{\textbf{Best classifier hyperparameters on engineered feature vectors}. Column ``Mean accuracy" gives the average accuracy of the classifier under a $K=20$-fold shuffle-split cross-validation using the engineered feature vectors in $\mathcal{C}_{1}$. Column ``Hero test" accuracy shows the accuracy when the classifier is trained using the entirety of $\mathcal{C}_{1}$ with the hyperparameters set to the specified value(s), and the classifier it learns is evaluated using $20900$ \emph{previously unseen} feature vectors.}
\label{tab:bestHyperparams}
\end{center}
\end{table*}

As noted in the main text, the performance of ML algorithms can depend quite strongly on their hyperparameters. Table \ref{tab:hyperparams} gives the values of the hyperparameters we considered in this work.

\begin{figure}
\includegraphics[width=\columnwidth]{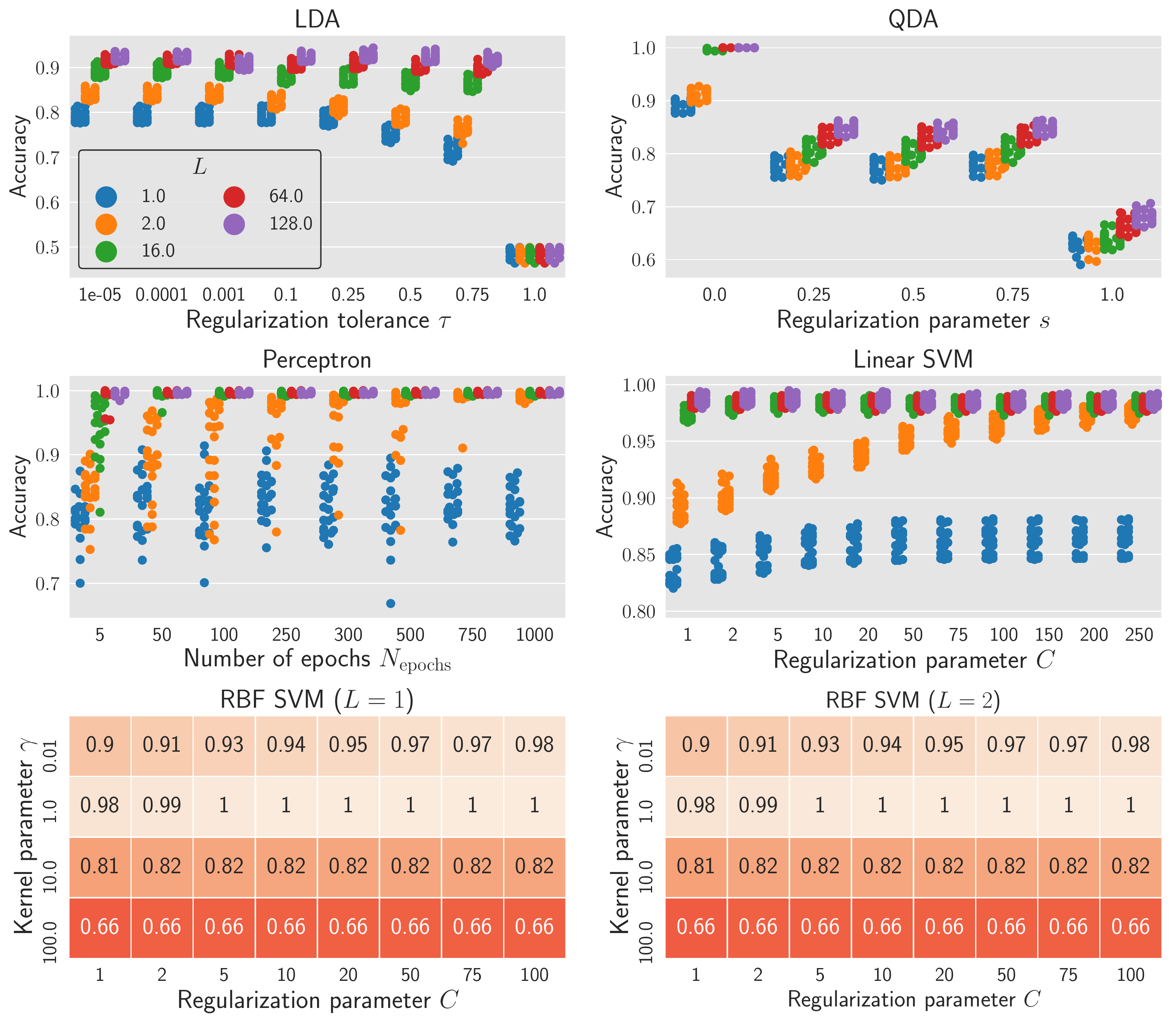}
  \caption{\textbf{Classification accuracy on $\mathcal{C}_{L}$ depends on the hyperparameter of the classification algorithm}. For the top 4 plots, the hue indicates the value of $L$. For the bottom 2, the \emph{average} classification accuracy is indicated in the heatmap.}
\label{fig:cv_hyperparams}
\end{figure}

\begin{figure}
\includegraphics[width=\columnwidth]{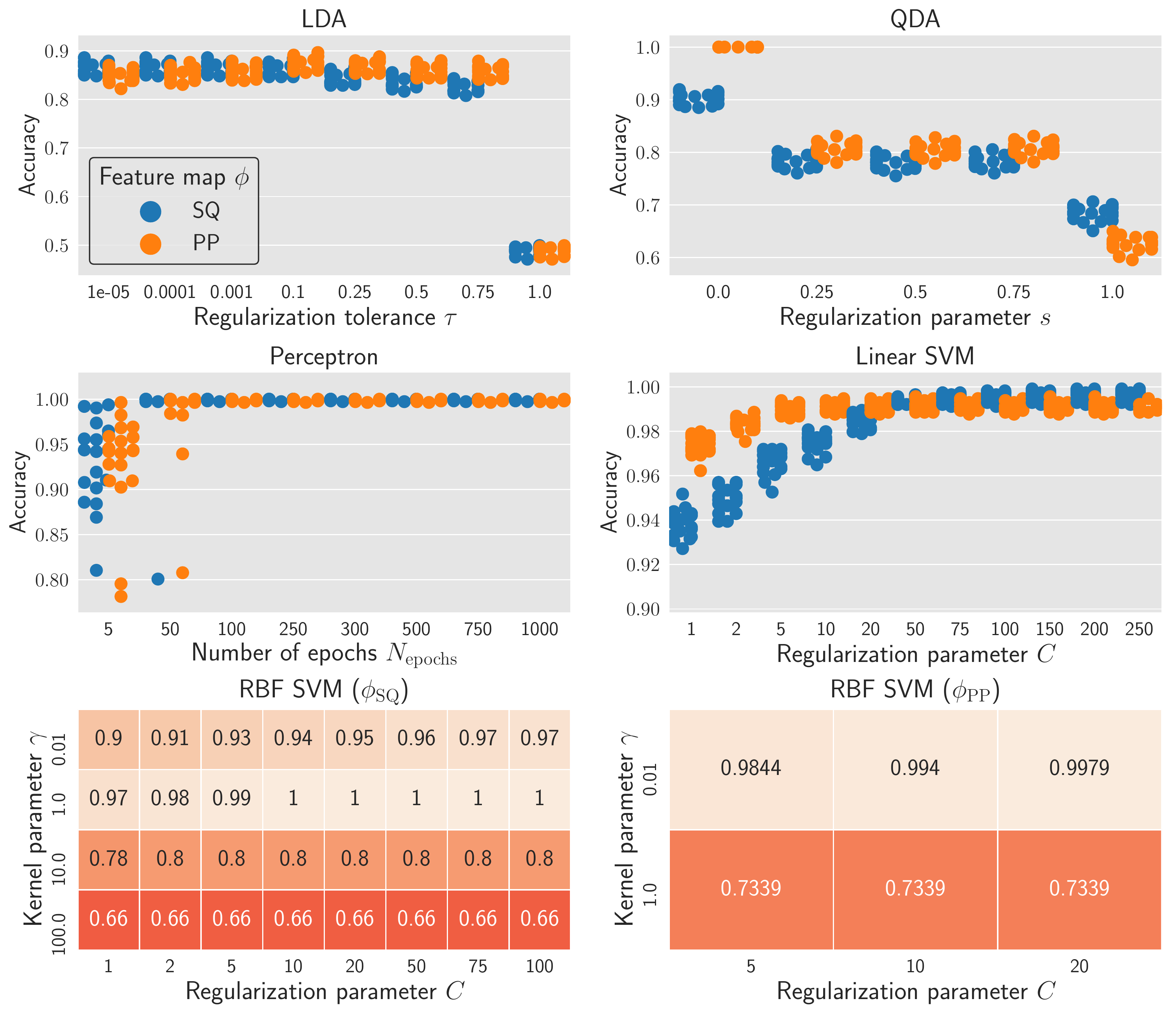}
 \caption{\textbf{Hyperparameter sweep on engineered feature vectors}.
For the top 4 plots, hue indicates which feature map was used. For the bottom 2, \emph{average} classification accuracy is indicated in the heatmap.}
\label{fig:cv_engineering}
\end{figure}

To determine which hyperparameter choices were best, we cross-validated the accuracy of the ML algorithms using 20 independently sampled training and testing sets, with 10\% of the data used for testing. Figure \ref{fig:cv_hyperparams} show the results of sweeping the hyperparameters and evaluating classification accuracy on $\mathcal{C}_{L}$. Figure \ref{fig:cv_engineering} shows results for the feature engineered feature vectors. Table \ref{tab:bestHyperparams} gives best hyperparameter values for learning using the engineered feature vectors, as determined by mean cross-validation accuracy (``Mean accuracy" column). We evaluated the chosen hyperparameters using a ``hero test": each algorithm was trained using all the feature-engineered vectors in $\mathcal{C}_{1}$ with the hyperparameters set to their optimal values. Then, the classifier learned by the algorithm was evaluated on $20900$ \emph{previously unseen} feature vectors. The accuracy is comparable to that evaluated under cross-validation, which suggests we have not inadvertently over-fit the training data.

~\newpage
\bibliographystyle{apsrev4-1}
\bibliography{GST_ML}

\end{document}